\documentclass[10pt,journal,compsoc]{IEEEtran}

\ifCLASSOPTIONcompsoc
\usepackage[nocompress]{cite}
\else
\usepackage{cite}
\fi
\ifCLASSINFOpdf
\else
\fi
\usepackage{amsmath,amssymb,amsfonts}
\usepackage{graphicx}
\usepackage{textcomp}
\usepackage{xcolor}

\usepackage{mathrsfs}

\usepackage{amsmath}
\usepackage{amsfonts}
\usepackage{amssymb}
\usepackage{multirow}
\usepackage{algorithm}
\usepackage{algpseudocode}
\usepackage{graphicx}
\usepackage[hyphens]{url}
\usepackage{amsthm}
\usepackage{subcaption}

\graphicspath{{figures/}}

\newtheorem{theorem}{Theorem}[section]
\newtheorem{lemma}{Lemma}[section]
\newtheorem{definition}{Definition}[section]

\newtheorem{example}{Example}[section]

\usepackage{color,soul}

\usepackage{tikz}

\usepackage{environ}
\usepackage{setspace} 

\newenvironment{alignSmall}{\nobreak\small\noindent\align}{\endalign}
\newenvironment{alignFootnotesize}{\nobreak\footnotesize\noindent\align}{\endalign}
\newenvironment{alignScriptsize}{\nobreak\scriptsize\noindent\align}{\endalign}

\NewEnviron{myAlignS}{%
	\begin{singlespace}
		\vspace{-3pt}
		\begin{alignSmall}
			\BODY
		\end{alignSmall}
		\vspace{-5pt}
	\end{singlespace}
}

\NewEnviron{myAlignSS}{%
	\begin{singlespace}
		\vspace{-3pt}
		\begin{alignFootnotesize}
			\BODY
		\end{alignFootnotesize}
		\vspace{-5pt}
	\end{singlespace}
}

\NewEnviron{myAlignSSS}{%
	\begin{singlespace}
		\vspace{-5pt}
		\begin{alignScriptsize}
			\BODY
		\end{alignScriptsize}
		\vspace{-5pt}
	\end{singlespace}
}

\newcommand {\geoind}{Geo-Indis\-tin\-guish\-ability}

\begin{document}

\title{Protecting Spatiotemporal Event Privacy in Continuous Location-Based Services}

%
%
%
%

\author{Yang~Cao,
	Yonghui~Xiao, 
	Li~Xiong,
	Liquan~Bai,
	and Masatoshi Yoshikawa
	\IEEEcompsocitemizethanks{
		\IEEEcompsocthanksitem Yang Cao and Masatoshi Yoshikawa are with the Department
		of Social Informatics, Kyoto University, Kyoto, Japan, 606-8501.\hfil\break 
		E-mail: \{yang, yoshikawa\}@i.kyoto-u.ac.jp.
		\IEEEcompsocthanksitem Yonghui Xiao is with Google Inc., Mountain View, CA, 94043\hfil\break 
		E-mail: yohu@google.com.
		\IEEEcompsocthanksitem Xiong Li and  Liquan Bai are with the Department
		of Computer Science, Emory University, Atlanta, GA, 30322.\hfil\break 
		E-mail: \{lxiong, liquan.bai\}@emory.edu.
	}
\thanks{Yang and Yonghui contributed equally to this work.}
\thanks{This paper is extended from \cite{cao_priste:_2019}.}
\thanks{Manuscript received XXX; revised XXX.}
}

%
%

\markboth{Journal of \LaTeX\ Class Files,~Vol.~XX, No.~X, August~201X}%
{Shell \MakeLowercase{\textit{et al.}}: Bare Demo of IEEEtran.cls for Computer Society Journals}
%

\IEEEtitleabstractindextext{%
\begin{abstract}
 Location privacy-preserving mechanisms (LPPMs) have been extensively studied for protecting users'  location privacy by releasing a perturbed location to  third parties such as location-based service providers.
 However, when a user's perturbed locations are  released continuously, existing LPPMs may not protect the sensitive information about  the user's spatiotemporal activities, such as ``visited hospital  in the last week" or ``regularly commuting between Address 1 and Address 2" (it is easy to infer that Addresses 1 and 2 may be home and office), which we call it \textit{spatiotemporal event}.
 In this paper, we first formally define  {spatiotemporal event}  as Boolean expressions between location and time predicates, and then we define  $ \epsilon $-\textit{spatiotemporal event privacy} by extending the notion of differential privacy.
 Second, to understand how much spatiotemporal event privacy that existing LPPMs can provide, we design computationally efficient  algorithms to quantify the privacy leakage of state-of-the-art LPPMs when an adversary has prior knowledge of the user's initial probability over possible locations.
It turns out that the existing LPPMs cannot adequately protect spatiotemporal event privacy.
 Third, we propose a framework, PriSTE, to transform an existing LPPM into one protecting spatiotemporal event privacy against adversaries with \textit{any} prior knowledge.
Our experiments on real-life  and synthetic data verified that the proposed method is effective and efficient.
\end{abstract}
	
	}

\maketitle

\section{Introduction}
The continued advances and usage of smartphones and GPS-enabled devices have provided tremendous opportunities for Location-Based Service (LBS),  such as Yelp or Uber for snapshot or continuous queries, for example, ``where is the nearest restaurant'' or ``continuously report the taxis within one mile of my location''.
Mobile users have to share their real-time location, or a sequence of locations with the service providers, which raises privacy concerns since users' digital trace can be used to infer sensitive information, such as home and workplace,  religious places and sexual inclinations \cite{golle_anonymity_2009} \cite{recabarren_what_2017}\cite{argyros_evaluating_2017}.

A large number of studies (see surveys  \cite{chatzikokolakis_methods_2017}\cite{liu_location_2018}\cite{primault_long_2018}) have explored how to protect user's location privacy which can be categorized from different aspects:  privacy goals, adversarial models, location privacy metrics, and location privacy preserving mechanisms (LPPMs).
\textit{Privacy goals} indicate what should be protected or what are the secrets (e.g., a  single location or a trajectory); \textit{adversarial models} make assumptions about the adversaries; \textit{location privacy metrics} formally define the quantitative measurement of the protection w.r.t. the privacy goal;  LPPMs is designed to achieve  a specified privacy metric.
For instance,  \geoind{} \cite{andres_geo-indistinguishability:_2013}  is a location privacty metrics, which is receiving increasing attention since it extends the notion of differential privacy \cite{dwork_differential_2008} to the location privacy setting so that the protection level does not depend on  adversaries' prior knowledge; the privacy goal of  \geoind{}  is to protect a single location (can be extended for protecting location trace \cite{chatzikokolakis_predictive_2014}); Laplace Planar Mechanism \cite{andres_geo-indistinguishability:_2013}  is  an LPPM satisfying  \geoind{}.
Another example is  Planar Isotropic Mechanism  \cite{xiao_protecting_2015}  for the metrics of $\delta$-location set privacy  to protect each location in a trajectory.
These state-of-the-art LPPMs take an actual location and a privacy parameter as inputs and probabilistically output a randomly perturbed location. 
A LPPM privacy parameter controls the location privacy level. 
For examples of the above mechanisms, the privacy parameter is deﬁned as a positive real value and a smaller privacy parameter indicates stronger privacy protection. 
In other words, the privacy parameter can be considered as the controlled level of privacy loss.

We argue that the existing state-of-the-art  LPPMs may not adequately protect users' sensitive information in their spatiotemporal activities because the  \textit{privacy goal}  is not well-studied.
The existing LPPMs focused on the protection of either a single location or a trajectory, which does not completely reflect the secrets that should be protected in users' spatiotemporal activities. 
To explain this, we need to define ``spatiotemporal activities".
We  define a  user's a single location at time $ t $ as a predicate $ l_t=s_i $  where   $ l_t $  is a variable representing  the user's position at time $ t $ and $ s_i\in \mathbb{S}, i\in[1,m]$ is a location on the map $ \mathbb{S} $ of $ m $ locations.
The value of such predicate can be either \textit{true} or \textit{false}, which could be a secret of the user.
Then, we can represent users' spatiotemporal activities as Boolean expressions of combining different predicates over spatial and/or temporal dimensions, which is called \textit{spatiotemporal event} in this paper.

\begin{figure}[t]
	\centering
	\includegraphics[width=8.5cm]{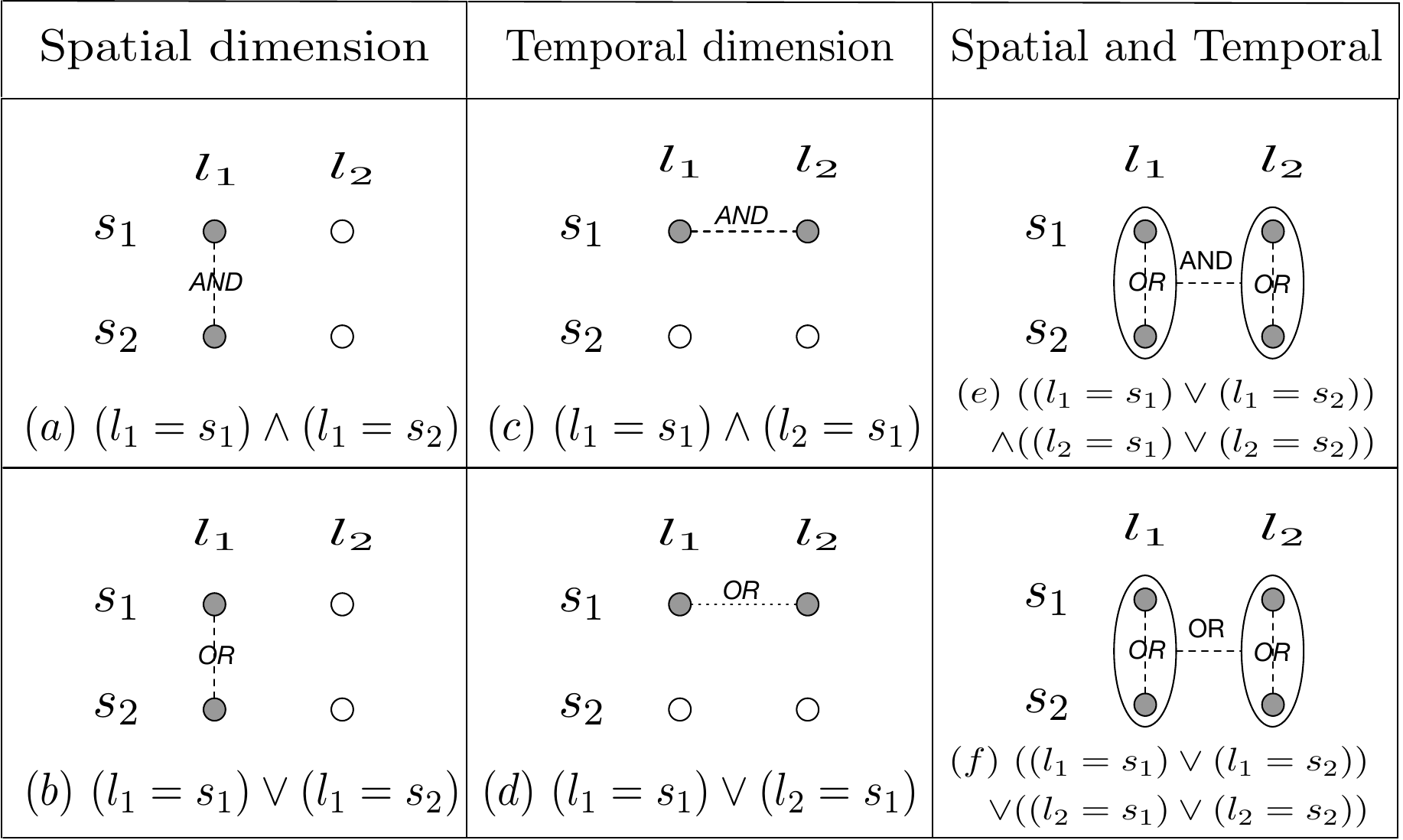}
	\caption{Six examples of spatiotemporal events. 
		Event (a) is always false. 
		Event (b) represents a sensitive \textbf{region}.
		Event (c)  represents a sensitive \textbf{trajectory}.
		Event (d)  represents the \textbf{presence} or not in a sensitive location.
		Event (e)  indicates  a mobility \textbf{pattern} passing	through sensitive regions. 
		Event (f)  indicates the \textbf{presence} or not in a sensitive region. 
	}
	\label{fig:stevent}
	\vspace{-12pt}
\end{figure}

In Fig.\ref{fig:stevent}, we illustrate  six representative Boolean expressions between location and time dimensions.
We use $ s_1$ and $s_2 $ to denote two locations on the map $\mathbb{S} $, and use $ l_1 $ and $ l_2 $ to denote two variables about a user's  locations at timestamps $ 1 $ and $ 2 $, respectively. 
Event (a) is always false since a user cannot  be physically at two different locations at the same time. 
Event (b) means that the secret is a sensitive region (or area) of $ \{s_1,s_2\} $. 
Event (c) represents a sensitive trajectory $ s_1 \rightarrow s_1 $ between timestamps 1 and 2, i.e., the user stays at $ s_1 $ at time 1 and time 2. 
Event (d) denotes that the secret is the visit to $ s_1 $ at timestamp $ 1 $ {or}  $  2$. 
Event (e) depicts the secret as a type of trajectory \textbf{pattern}, i.e., the user may stay at two sensitive regions successively; a real-world example of such event is ``regularly commuting between Address 1 and Address 2 every morning and every afternoon", i.e.,  periodic spatiotemporal events may happen every week day.
Event (f) indicates the secret as user's {\textbf{presence}} in sensitive region $ \{s_1,s_2\} $ at  either timestamp  $ 1 $ or  $ 2 $; a real-world example of such event is ``visited hospital  in the last week", i.e., the hospital visit may happen once or multiple times at any time in last week.

We can see that the spatiotemporal events representing sensitive locations and a trajectory (i.e., (b) and (c)), which are the  major privacy goals of  existing LPPMs,  are only two cases  among the six  enumerated examples.
Hence, even if an LPPM protects each location or a trajectory, it may not protect a complex spatiotemporal event such as the ones shown in Fig.\ref{fig:stevent} (e) and (f) since such new  privacy goals have not been formalized in the literature.

 There are three challenges in protecting such a new privacy goal.
 First, we lack the formal definition of spatiotemporal event and privacy metrics for it.
 Second, evaluating the privacy guarantee of a given spatiotemporal event could be computationally intractable since the  event can be  extremely complicated.
 Taking the \textbf{pattern} event (e.g., Fig.\ref{fig:secrets_pairs}(f)) for example, if the sensitive region includes $ m $ locations and the length of such event spans $ T $ timestamps, there are $ m^T $ possible trajectories that need to be protected, which may lead to exponential  time computation.
 Third, similar to \geoind, we hope to design a  mechanism that is robust to adversaries with \textit{any} prior knowledge.

\vspace{2mm}\noindent
{\bf Contributions.}
To the best of our knowledge, this is the first paper that studies how to protect spatiotemporal event privacy.
Our contributions  are summarized as follows.

First, we study the privacy goal and privacy metric for protecting spatiotemporal event (Section \ref{sec:def}).
We formally define a new type of privacy goal, i.e., spatiotemporal events, as Boolean expressions of  location-time predicates, and propose a privacy metric, \textit{$ \epsilon $-spatiotemporal event privacy}, for protecting spatiotemporal events by extending  the notion of differential privacy.
We also explore the difference between the metrics of location privacy and spatiotemporal event privacy.
It turns out that, although the definition of spatiotemporal event  is more general than  a single location or a trajectory, the privacy metrics between spatiotemporal event privacy and location privacy can be orthogonal.
Hence, it would be preferable that an LPPM achieving a location privacy metric such as Geo-indistinguishability can also satisfy $ \epsilon $-spatiotemporal event privacy w.r.t. user-specified events.
Location privacy  provides general protection against unknown risks, while spatiotemporal event privacy guarantees flexible and customizable protection which may not be provided by the existing LPPMs.

Second, we develop efficient algorithms  for quantifying how much $ \epsilon $-spatiotemporal event privacy  a given LPPM can provide w.r.t. adversaries with a specific prior knowledge about the user's  initial probability distribution over possible locations  (Section \ref{sec: quantifying}).
We model  an LPPM as an emission matrix that takes user's true position as input and outputs a perturbed location.
As we mentioned previously, one of the challenges in quantifying the probability of a  spatiotemporal event is that the computational complexity may be exponentially increasing with the number of predicates in a user-specified spatiotemporal event.
We develop a novel \textit{two-possible-world} method to quantify spatiotemporal event privacy with  linear complexity.

Third, based on our quantification method, we propose a framework, i.e, PriSTE (\underline{Pri}vate \underline{S}patio-\underline{T}emporal \underline{E}vent), which converts a mechanism for location privacy  into one for spatiotemporal event privacy against adversaries with any prior knowledge (Section \ref{sec:priste}).
 We demonstrate the effectiveness of our framework by two case studies using state-of-the-art  LPPMs, i.e., Laplace Planar Mechanism for Geo-indistinguishability \cite{andres_geo-indistinguishability:_2013}  and Planar Isotropic Mechanism
 for $\delta$-location set privacy \cite{xiao_protecting_2015}.




Finally, we  evaluate our algorithms on both synthetic and real-world datasets testing its feasibility, efficiency, and the impact of various parameters (Section \ref{sec:exp}).


\begin{table*}[!]
	\scriptsize
	\centering
	\begin{tabular}{|c|c|c|}
		\hline
		{\centering {\bf \textbf{Spatiotemporal Event}}}&{\centering {\bf Boolean Expression} }&{\centering {\bf Interpretation (if the event is true)} }\\
		\hline
		single location event&$l_t=s_{i}$& visited the location $s_{i}$ at time $t$  \\\hline
		$\textsc{Presence}$ at a single location &$(l_1=s_{i})\lor (l_2=s_{i})\lor \cdots \lor (l_T=s_{i})$& visited  a location $s_{i}$ during time $\{1,2,\cdots,T\}$ \\\hline
		single region  event&$(l_t=s_{i})\lor (l_t=s_{j})\lor \cdots \lor (l_t=s_{k})$ &  visited a region $\textbf{s}=\{s_{i} ,\cdots, s_{k}\}$ at time $t$\\\hline
		single trajectory event&$(l_1=s_{i})\land (l_2=s_{j})\land \cdots \land (l_T=s_{k})$& visited specified \textit{locations} successively during  time $\{1,2,\cdots,T\}$  \\\hline
		$\textsc{Pattern}$ of trajectories & \begin{tabular}[l]{@{}c@{}}$((l_1=s_{i})\lor (l_1=s_{j})\lor\cdots \lor (l_1=s_{k}))\land\cdots $\\$\land((l_T=s_{l})\lor (l_T=s_{m})\lor\cdots \lor  (l_T=s_{n}))$\end{tabular}& visited specified \textit{regions} successively during  time $\{1,2,\cdots,T\}$ \\\hline
	\end{tabular}
	\caption{Representative  examples of $\textsc{Events}$ of Boolean operations on the (location, time) predicates.}
	\label{tbl-events}
\end{table*}

\section{Defining Spatiotemporal Event Privacy}
\label{sec:def}

\subsection{Scenario}
\label{subsec:problem-statement}
We consider a scenario that a single user continuously releases her perturbed location  with an untrusted third party such as location-based service provider.  
The user's true locations are denoted by {\footnotesize $l_1,l_2,\cdots,l_T$}.
A location privacy-preserving mechanism (LPPM) blurs user's true location $  l_t$ to a perturbed one $  o_t$ that satisfies a privacy metric such as \textit{\geoind{}}\cite{andres_geo-indistinguishability:_2013} or \textit{$\delta$-location set privacy} \cite{xiao_protecting_2015}.
Essentially, the LPPM is an emission matrix that takes user's true location as input and outputs a perturbed one.
The major notations in this paper are summarized in Table \ref{tbl-denotation}.
\renewcommand{\arraystretch}{1.2}
\begin{table}[!htbp]
	\small
	\begin{tabular}{|p{25pt}|p{200pt}|}
		\hline
		$m$ & the amount of all possible locations on the map \\\hline
		$\textbf{s}$ & a  vector representing a region, $\textbf{s} \in \{0,1\}^{m\times 1}$ \\\hline
		$t$ &a timestamp in $\{1,2,\cdots, T\}$\\\hline
		$\mathcal{S}$ & a  sequence of  regions \\\hline
		$\mathcal{T}$ & a sequence of timestamps  \\\hline
		$l_t$ & a user's true location at time $t$\\ \hline
		$o_t$ & a user's perturbed location at time $t$\\ \hline
		$\textsc{Event}$ & a spatiotemporal event\\\hline
		$\tilde{\textbf{p}}_{o_t}$ & a vector of emission probabilities given the observation $o_t$.\\\hline
		$\tilde{\textbf{p}}_{o_t}^\textbf{D}$ &  a diagonal matrix with the vector $\tilde{\textbf{p}}_{o_t}$ on the diagonal. \\\hline
	\end{tabular}
	\caption{Notations.}
	\label{tbl-denotation}
\end{table}

\subsection{Spatiotemporal Events}

We fist define {location-time predicate}, which is an atomic element in spatiotemporal events.
Let $\mathbb{S}=\{s_1,\cdots, s_m\}$  be the domain of all possible locations, where $m$ is the size of the  domain and $ s_i $ is one location (we use \textit{state} interchangeably) on the map.
At time  $ t $, a user's location  can be stated as $  l_t=s_{i}$, which means the user is at location $ s_i $ at time $ t $.
We call $  l_t=s_{i}$ \textit{location-time predicate}, whose value can be \textit{true} or \textit{false} depending on the ground truth of user's state at $ t $.


We define spatiotemporal events as Boolean expressions of the location-time predicates. 

\begin{definition}[$\textsc{Event}$]
	A spatiotemporal event, denoted by $\textsc{Event}$, is a single location-time predicate or a combination of location-time predicates linked by the Boolean operators AND, OR, NOT (i.e., $\land$, $\lor$, $\neg$, respectively).
\end{definition}
%
Using Boolean logic to define spatiotemporal events enables  users to customize their privacy preference for diverse real-world activities.
Table \ref{tbl-events} shows some representative examples of $\textsc{Event}$. 
For example, an event of a single location can be represented by  a predicate alone, i.e.,  $l_t=s_i$;
a single trajectory event can be denoted by $(l_1=s_{i})\land (l_2=s_{j})\land \cdots,\land (l_T=s_{k})$, which is true if the user passes through $ s_i, s_j, \cdots, s_k $ during time $ 1$  to $ T $.
%

For the ease of exposition, we define the following notations.
We denote a region  (i.e., a set of locations)  by a vector $\textbf{s}\in \{0,1\}^{m\times 1}$ where the $i$th element is $1$ only if  the region contains $s_i$. 
We use $ \mathcal{S} $ to indicate a sequence of regions.
We denote the corresponding timestamp of each region by $\mathcal{T}$ as a sequence of timestamps with the same  cardinality  of $ \mathcal{S} $.
A pair of  i-th elements in $  \mathcal{S} $ and $ \mathcal{T} $ could form  a \textit{single region event} as shown in Table \ref{tbl-events} (or Event (b) in Fig. \ref{fig:stevent}).
These single region events could be  combined by AND or OR, which form  \textsc{Presence} or \textsc{Pattern} (see the difference between Events (e) and (f) in Fig. \ref{fig:stevent}).

\subsubsection{\textsc{Presence} Event}
When the secret is whether or not a user visited a sensitive region (e.g., medical	facilities) in a given time period,
we can use \textsc{Presence} to represent such secret.
A \textsc{Presence} event holds if a user appears in \textit{any} one of the regions with user-specified timestamps. 
In the simplest case of  \textsc{Presence}, when the region includes only one location and the time period consists of one timestamp,  it reduces to a \textit{single location event} shown in Table \ref{tbl-events}.
Hence, \textsc{Presence} event can be seen as a generalization of single location event.


\begin{definition}[\textsc{Presence}]
	\label{def-presence}	
Given a sequence of regions $\mathcal{S}=[\textbf{s}_1,\cdots,\textbf{s}_n]$ and a sequence of timestamps $\mathcal{T}=[t_1,\cdots,t_n]$, 
if a user appears in at least one $\textbf{s}_k \in\mathcal{S}$ at the corresponding time $ t_k \in\mathcal{T}$,then it is a  presence event, denoted by
$\textsc{Presence}(\mathcal{S},\mathcal{T})$.
\end{definition}

\begin{figure}[t]
	\centering
	\includegraphics[width=7.5cm]{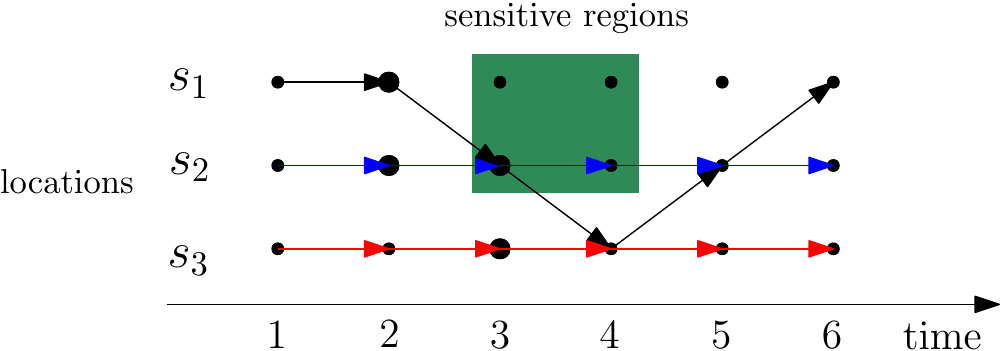}
	\caption{We show two events, i.e., \textsc{Presence}($ \mathcal{S}, \mathcal{T} $)  and \textsc{Pattern}($ \mathcal{S}, \mathcal{T} $). If the user's ground truth trajectory is the black one, only \textsc{Presence}($ \mathcal{S}, \mathcal{T} $) is true; if the user's trajectory is the blue one, both events are true; if the user's trajectory is the red one, both events are false.}
	\label{Figure-MM-example1}
	\vspace{-5pt}
\end{figure}

\begin{example}[Example of \textsc{Presence}]
	\label{ex1}
	Figure \ref{Figure-MM-example1} shows a map of $\mathbb{S}=\{s_{1},s_{2},s_{3}\}$. 
	For this event, the region $\textbf{s}=[1,1,0]^{\intercal}$ denoting the states $s_{1}$ and $s_{2}$; the time period $\mathcal{T}=[3,4]$ denoting timestamp $3$ and $4$. 
	Let $ \mathcal{S} = [\textbf{s}, \textbf{s} ] $.
	This \textsc{Presence} $ (\mathcal{S}, \mathcal{T}) $ event is expressed as $(l_3=s_{1})\lor(l_3=s_{2})\lor(l_4=s_{1})\lor(l_4=s_{2})$.
	The shaded region shows a \textsc{Presence} event that the user appears in a region of $ \{ s_1, s_2\} $ during timestamps $3$ and $4$. 
	If the user's true trajectory passes through the shaded region (at least one timestamp), the event is true.
\end{example}

	\vspace{-5pt}
\subsubsection{\textsc{Pattern} Event}
We  use \textsc{Pattern} to represent the secret  whether or not a user visited multiple sensitive regions successively.
In a simple case of \textsc{Pattern}  event, the regions consist of single locations at a sequence of timestamps, then it is reduced to \textit{single trajectory event} shown in Table \ref{tbl-events}.
Hence, \textsc{Pattern} is a generalization of a user's trajectories.
\begin{definition}[\textsc{Pattern}]
	\label{def-pattern}
	Given a sequence of regions $\mathcal{S}=[\textbf{s}_1,\cdots,\textbf{s}_n]$ and a sequence of timestamps $\mathcal{T}=[t_1,\cdots,t_n]$, if a user appears in all $\{\textbf{s}_1,\cdots,\textbf{s}_n\}$ sequentially at the corresponding time  during $\mathcal{T}$, 
	then it is a pattern event, denoted by $\textsc{Pattern}(\mathcal{S},\mathcal{T})$.
\end{definition}
\begin{example}[Example of \textsc{Pattern}]
	The \textsc{Pattern} event in Fig. \ref{Figure-MM-example1} represents  trajectories with a pattern going through a sensitive region {\footnotesize $\{s_{1},s_{2}\}$} at timestamp $2$ and  the same region {\footnotesize $\{s_{1},s_2\}$} at timestamp $3$ successively. 
	This \textsc{Pattern} event is expressed as {\footnotesize $((l_2=s_{1})\lor(l_2=s_{2}))\land((l_3=s_{1})\lor(l_3=s_{2}))$}.
\end{example}

\vspace{-9pt}
\subsubsection{Remarks}
From the above definitions, we can see that, in terms of privacy goal, spatiotemporal event privacy can be a \textit{generalization} of location privacy studied in the literature in which the privacy goal is protecting a single location or a trajectory.
In this paper, we focus on the two representative  events defined above, i.e., \textsc{Presence} and \textsc{Pattern}, which are the two most complicated and unexplored  events among examples in Fig.\ref{fig:stevent}. 
We note that $\textsc{Presence}$ and $\textsc{Pattern}$  include the cases when the time $\mathcal{T}$ is not consecutive.
Users can specify one or multiple  spatiotemporal events to be protected.
We propose  a privacy metric for preserving user's indistinguishability of her specified spatiotemporal events in the next section.

\vspace{-5pt}
\subsection{$ \epsilon $-Spatiotemporal Event Privacy}

Inspired by the definition of differential privacy\cite{dwork_differential_2008}, we define $ \epsilon $-Spatiotemporal Event Privacy as follows.
\begin{definition}[$\epsilon$-Spatiotemporal Event Privacy]
	\label{def:e-STE}
	A mechanism preserves $\epsilon$-Spatiotemporal Event Privacy for a spatiotemporal \textsc{Event} if at any timestamp $t$ in $\{1, \cdots,T\}$ given any observations $\{o_1,\cdots,o_t\}$,
	\begin{myAlignS}
	\label{eqn-eps-delta-DP1}
	\Pr(o_1,\cdots,o_t|\textsc{Event}) \leq e^{\epsilon}\Pr(o_1,\cdots,o_t| \lnot\textsc{Event})
	\end{myAlignS}
	\vspace{-8pt}
	\noindent	where  {\textsc{Event}} is a logic variable about the user-specified spatiotemporal event and  $\lnot\textsc{Event}$ denotes the negation of $\textsc{Event}$. {\small $ \Pr(o_1,o_{2},\cdots,o_t|\textsc{Event}) $} denotes the probability of the observations $ o_1,o_{2},\cdots,o_t $ given the value of \textsc{Event}. 
\end{definition}

There are two major benefits of extending differential privacy to protecting spatiotemporal events.
First, it provides a well-defined semantics for spatiotemporal event privacy. 
Similar to differential privacy that requires the indistinguishability between any two neighboring databases\cite{dwork_differential_2008}, $\epsilon$-Spatiotemporal Event Privacy requires the indistinguishability regarding whether  the $\textsc{Event}$ is true or false given any observations.
 It provides a clear privacy semantics: it is hard for adversaries to distinguish  whether the event happened or not. 
Another benefit is that, similar to differential privacy whose privacy  guarantee is independent of the prior probability of a given database,  the privacy provided by $\epsilon$-Spatiotemporal Event Privacy  is independent of the prior probability of the protected event.

\begin{figure}[t]
	\centering
	\includegraphics[width=9cm]{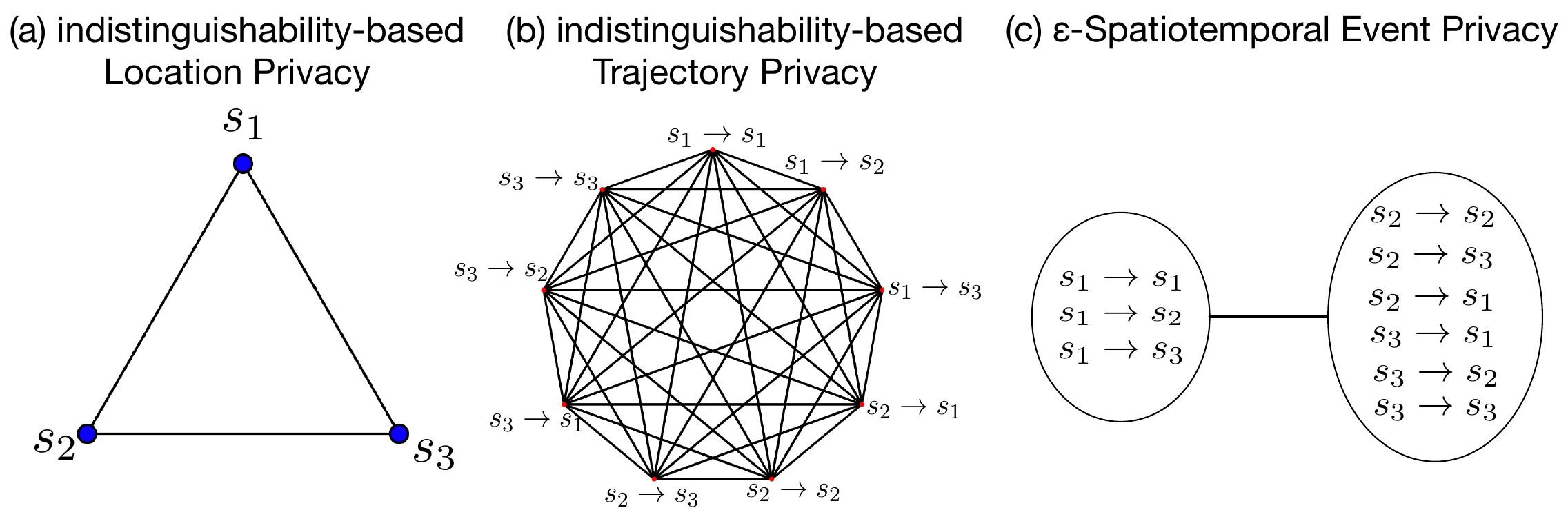}
	\caption{Illustration of  indistinguishability-based privacy metrics for distinct privacy goals when {\footnotesize $ {\mathcal{S}}=\{s_1,s_2,s_3\}$} and {\footnotesize $  \mathcal{T}=2$}.}
	\label{fig:secrets_pairs}
\end{figure}

To better understand  the characteristics of spatiotemporal event privacy, we illustrate the indistinguishability-based privacy metrics for the three privacy goals in Fig.\ref{fig:secrets_pairs}, where the lines connecting two secrets indicate the requirements of indistinguishability between the corresponding two possible values of the secrets.

As shown in Fig.\ref{fig:secrets_pairs} (a), indistinguishability-based location privacy metrics (such as geo-indistinguishability\cite{andres_geo-indistinguishability:_2013}) require indistinguishability  between each pair of locations.
Indistinguishability-based trajectory privacy metrics \cite{chatzikokolakis_predictive_2014}\cite{xiao_protecting_2015} \cite{theodorakopoulos_prolonging_2014} requires indistinguishability between each pair of possible trajectories as shown in Fig.\ref{fig:secrets_pairs} (b).
Whereas, $ \epsilon $-spatiotemporal event privacy  requires  indistinguishability between the defined event and its negation.
For example, if the spatiotemporal event is defined as $\textsc{Pattern}(\mathcal{S},\mathcal{T})$ where $  \mathcal{S}=[\textbf{s}_1, \textbf{s}_2], \textbf{s}_1=\{s_1\}, \textbf{s}_2=\{s_1, s_2, s_3\}$ and $ \mathcal{T}=[1,2] $ (i.e., a trajectory passes through $ s_1 $ and  then a region $ \{s_1, s_2, s_3\} $ successively), then it only requires the indistinguishability between the set of all possible trajectories that pass through $ \{s_1\} $ and $ \{s_1, s_2, s_3\} $ and the set of trajectories that do not. 
This  spatiotemporal event privacy makes sense when some mobility patterns are sensitive. 
For example, if $ s_1$ is ``hospital'', $ s_2 $ is ``home'',  and $ s_3$ is ``office'', the pattern from $ s_1 $ to  $\{s_1, s_2, s_3\} $ could be sensitive.

We note that  spatiotemporal event privacy is orthogonal to location privacy or trajectory privacy.
First, protecting the privacy of a single location or a trajectory  may not  imply the protection of spatiotemporal event  privacy because spatiotemporal event can be complex as shown in Fig.\ref{fig:stevent}(e) or (f).
The existing LPPMs  are  designed to ensure privacy metrics  defined on locations or trajectories. 
One of our focus in this study is to quantify  how much spatiotemporal event privacy  a given LPPM can provide, which will be elaborated in the next section.
Second,  protecting spatiotemporal event privacy does not  imply the protection of location privacy because they define  indistinguishability over different level of secrets.
Taking Fig. \ref{fig:secrets_pairs}(c) for example, the indistinguishability between $ s_1 \rightarrow s_1 $ and $ s_1 \rightarrow s_2 $ is not required in such spatiotemporal event privacy guarantee; however, it is required in trajectory privacy as shown in Fig. \ref{fig:secrets_pairs}(b).
Even if we define the event in spatiotemporal event privacy as a single location, say $ s_1 $, the  guarantee of  spatiotemporal event privacy is  the indistinguishability between $s_1 $ and $ \{s_2,s_3\}  $, which does not guarantee the indistinguishability between $ s_1 $ and $ s_2 $.

%

 It would be preferable if we achieve both location privacy and spatiotemporal event privacy so that a user can enjoy the best of two worlds: Location privacy  provides general protection against unknown risks when sharing location with the third parties, while spatiotemporal event privacy guarantees customizable protection which may prevent against profiling attacks \cite{recabarren_what_2017} \cite{de_mulder_identification_2008}.
Therefore, in this paper, we study how to use an existing probabilistic LPPM  (e.g., Laplace Planar Mechanism \cite{andres_geo-indistinguishability:_2013} and  Planar Isotropic Mechanism  \cite{xiao_protecting_2015}) to achieve  $ \epsilon $-spatiotemporal event privacy.

We note that the definition of events may reveal a user's sensitive information. 
In this paper, we assume that the events and the protection mechanisms are locally and securely stored in the user's device.
The user may specify one or multiple events that need to be protected.
In practice, we can also have default that are suggested by a privacy preference recommendation system for users' selection \cite{asada_when_2019} or pre-specified event templates that are given by the user.


\begin{figure}[t]
	\centering
	\includegraphics[width=8.8cm]{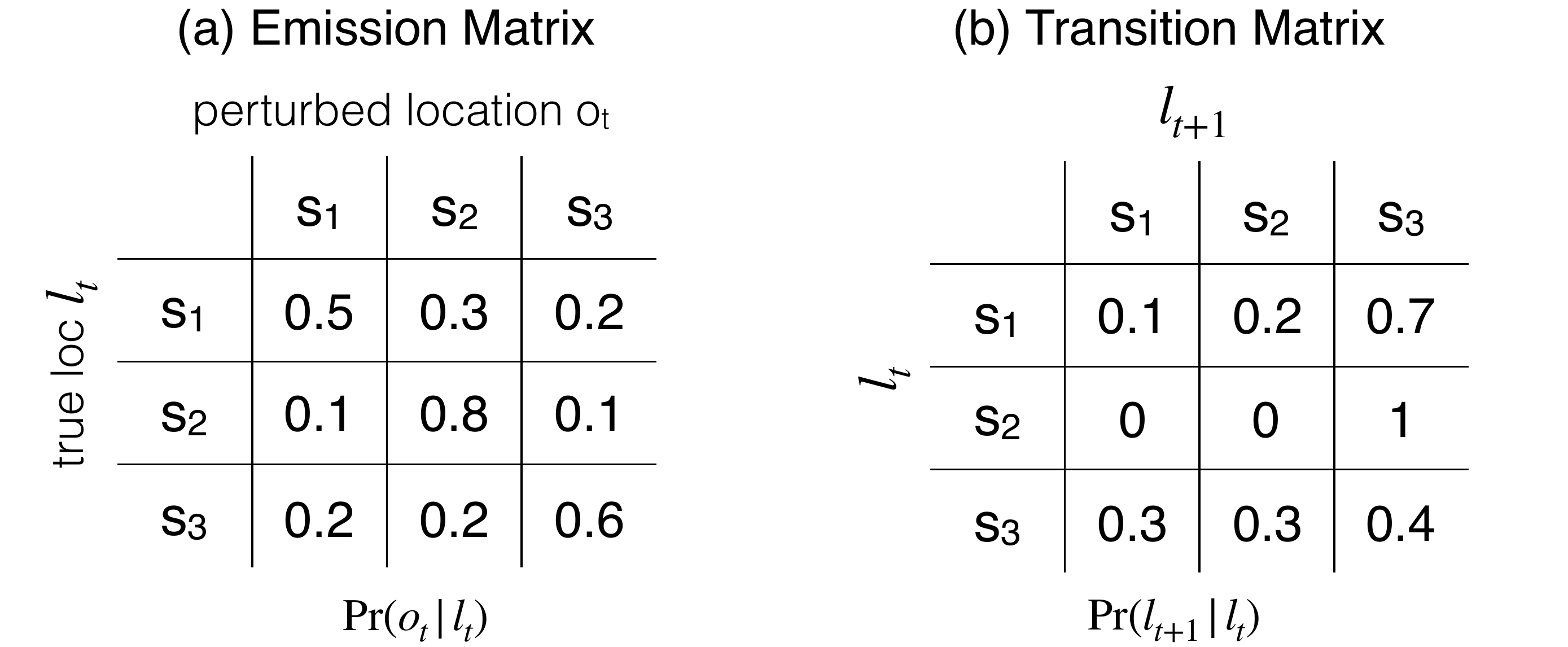}
	\caption{Illustration of  emission matrix and transition matrix.}
	\label{fig:tm-em}
\end{figure}

\section{Quantifying  Spatiotemporal Event Privacy}
\label{sec: quantifying}


\subsection{Overview of our approach} 
We first clarify our assumptions about LPPM, transition matrix and  user's initial probability distribution of each location as follows.
First, we consider an LPPM that takes input as user's true location $ l_t  $ and outputs a perturbed location $ o_t $ at time $ t $.
We use  an $ m \times m $ emission matrix  (as shown in Fig.\ref{fig:tm-em}(a)) to represent the LPPM.
Second, we assume a user's  location at time $ t+1 $ are correlated with her location at time $ t$,  representing by an $ m \times m $ transition matrix as shown in Fig.\ref{fig:tm-em}(b), and such transition matrix  is public information which can be  learned from either historical trajectory or the pattern of road networks. 
 We model the correlation between user's  consecutive locations using first-order\footnote{If the Markov model is high-ordered, i.e., the transition matrix has a larger state domain, our approach still works.}  time-homogeneous\footnote{If the Markov model is  time-varying, i.e., transition matrices at different $ t $ are not identical, our approach still works. We explain this in the next section.} Markov model, i.e., the transition matrix is identical at each $ t $.
The transition matrix is given in our system.
Third, for ease of exposition,  we first assume that adversaries who have a specific knowledge of the user's initial probability distribution over possible locations, which is denoted by $ \boldsymbol{\pi} $; in the next section, we will remove this assumption so that the  spatiotemporal event  privacy leakage will be bounded in $ \epsilon $ w.r.t. adversaries with any prior knowledge of user's initial probability.

Now, we explain the main goal of quantifying the spatiotemporal event privacy leakage  of the LPPM and our approach.
Based on Definition \ref{def:e-STE}  of $ \epsilon $-spatiotemporal event privacy,  we need  to calculate the maximum ratio of $\frac{\Pr(o_1,o_2,\cdots,o_T|\textsc{Event})}{\Pr(o_1,o_2,\cdots,o_T|\lnot\textsc{Event})}$  in which  $ o_1,o_2, \cdots,o_T $ are released by a given LPPM.
This ratio can be considered as spatiotemporal event privacy leakage w.r.t. the user-specified event. 
 We quantify this ratio w.r.t. given observations $ o_1,o_2, \cdots,o_T $ and a given user's initial probability $ \boldsymbol{\pi} $, so that we can directly calculate the $\Pr(o_1,o_2,\cdots,o_T|\textsc{Event}) $.
In Section \ref{sec:priste}, we will design a mechanism for  spatiotemporal event privacy w.r.t. any observations and arbitrary initial probability.
Our goal in this section is to calculate the likelihood of the observations given {\footnotesize $ \textsc{Event}  $} or {\footnotesize $ \lnot\textsc{Event}  $} , i.e., {\footnotesize $\Pr(o_1,o_2,\cdots,o_T| \textsc{Event}) $} or {\footnotesize $\Pr(o_1,o_2,\cdots,o_T| \lnot \textsc{Event}) $}, which can be derived by {\footnotesize $\Pr(o_1,o_2,\cdots,o_T|\textsc{Event}) = \frac{\Pr(o_1,o_2,\cdots,o_T, \textsc{Event}) }{\Pr( \textsc{Event}) } $}.
We call {\small $\Pr( \textsc{Event})  $} as \textit{prior probability} of the event, and {\small $\Pr(o_1,o_2,\cdots,o_T, \textsc{Event}) $} as \textit{joint probability} of the event.


A severe challenge of calculating the prior or joint probabilities of the event is the computational complexity.
Given an arbitrary spatiotemporal event, we need to enumerate all possible combination of the Boolean expression for prior and joint probabilities, which can be exponential to the number of predicates in the expression.
To address this problem, 
we propose a two-possible-world method for computing the prior and joint probabilities in Sections \ref{subsec:prior} and \ref{subsec:joint}.

For ease of exposition, we define notations below.
{\small $\textbf{M}\in\mathbb{R}^{m\times m}$} denotes a transition matrix that describes  temporal correlations in user's location.
At timestamp $1$, an initial probability is denoted by {\small $\boldsymbol{\pi}\in[0,1]^{1\times m}$}. During timestamp {\small $\{1, 2, \cdots, T\}$}, the probability of the true location {\small $\Pr(l_t)$} is denoted by a row vector {\small $\textbf{p}_t\in[0,1]^{1\times m}$} where the $i$th element denotes $\Pr(l_t=s_i)$. A Markov model follows the transition  property of {\small $\textbf{p}_{t+1}=\textbf{p}_t\textbf{M}$}, e.g., after a Markov transition, {\small $\textbf{p}_{2}=\boldsymbol{\pi}\textbf{M}$}  at timestamp $2$ given {\small $\textbf{p}_1=\boldsymbol{\pi}$}. 



The notations below for matrix computation are also used in the rest of this paper. 
Let $\textbf{0}$ and $\textbf{1}$ be  row vectors with $m$ elements being $0$ and $1$ respectively. 
$[\textbf{0},\textbf{1}]$ is a row vector in {\small $\mathbb{R}^{1\times 2m}$}. 
$\textbf{a}\circ\textbf{b}$ denotes the Hadamard product of $\textbf{a}$ and $\textbf{b}$. 
$\textbf{a}^\textbf{D}$ is a diagonal matrix with the elements of vector $\textbf{a}$ on the diagonal. 

\subsection{Computing Prior Probability of an Event}
\label{subsec:prior}
To avoid the exponential complexity, we propose an efficient algorithm with two possible worlds.
The idea is to elaborate a ``new" transition matrix $ \textbf{M}_t \in \mathbb{R}^{2m\times 2m}$ at each time $ t $ which encodes the complex spatiotemporal event inside, so that the calculation of the prior or joint probability for a complicated event is the same as one simple predicate.

\noindent{\bf \textbf{Intuition.}} 
The main idea of our method is to use two virtual worlds denoting whether the \textsc{Event} is true or false.  
The states in the two worlds denote the joint probabilities $\Pr(l_t=s_i,\textsc{Event})$ and $\Pr(l_t=s_i,\lnot\textsc{Event})$. 
For \textsc{Presence}, once a trajectory enters into the region of the \textsc{Presence}, its probability will be kept in the world of true \textsc{Event} forever. 
For \textsc{Pattern}, the probability distribution among the two worlds are derived at the beginning timestamp of the \textsc{Event}, and only the trajectories satisfying the \textsc{Pattern} will be kept in the world of true \textsc{Event}. At last, the sum of probabilities in the world of true \textsc{Event} will be $\Pr(\textsc{Event}\textrm{ is true})$.

In the following, we study how to compute the prior probabilities of \textsc{Presence} and 
\textsc{Pattern} events.
For simplicity, the events in the rest of the paper are defined in consecutive time and use $ start $ and $ end $ to denote the start time point and end time point of the user-specified spatiotemporal event. 
We  assume $\mathbb{S}$ to be $\{s_{1},s_{2},s_{3}\}$ in the following examples.

\subsubsection{Presence Events}

\begin{figure}[h]
	\centering
	\includegraphics[width=5cm]{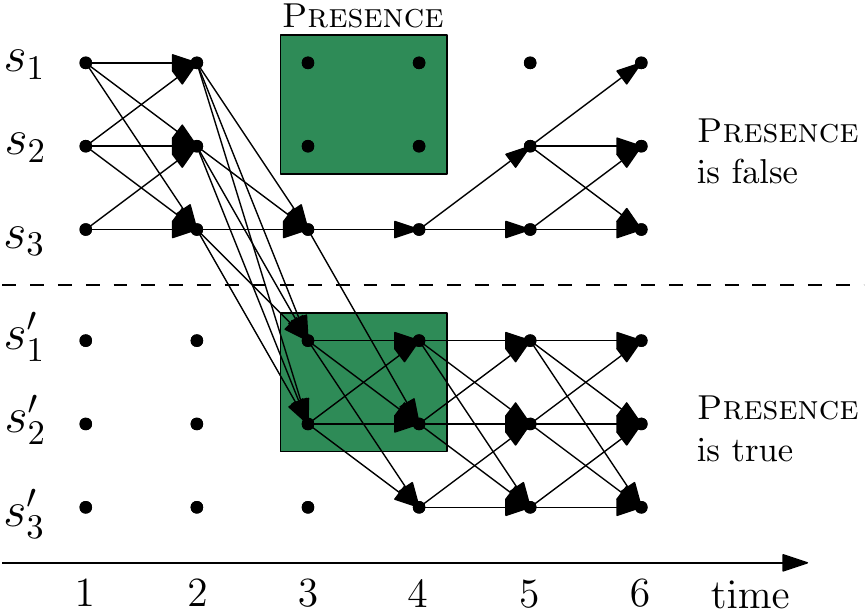}
	\caption{{\small New Markov transitions: all transitions going to the \textsc{Presence} region will be re-directed to the virtual worlds.}}
	\label{Figure-example-prior-presence}
\end{figure}

\begin{example}
	\label{example-prior-presence}
	Let us consider the same \textsc{Presence} event  defined in Example \ref{ex1}, 
	It is defined as an event passing through $s_1$ or $s_2$ during  $t=3$ or $t=4$, i.e., $\textbf{s}=[1,1,0]^{\intercal}$, $start=3, end=4$.
	The transition matrix $\textbf{M}$ is given below. 
	\begin{myAlignSSS}
	\textbf{M}=\left[
	\begin{scriptsize}
	\begin{array}{ccc}
	0.1 &0.2&0.7\\
	0.4 & 0.1&0.5\\
	0&0.1&0.9\\
	\end{array}
	\end{scriptsize}
	\right]
	\end{myAlignSSS}
	Then Fig.\ref{Figure-example-prior-presence}  shows the new transitions in the two worlds, the top world and the bottom world separated by the dashed line in Fig.\ref{Figure-example-prior-presence}, corresponding to the two possible worlds where the presence event is false or true respectively.  
	From time $1$ to $2$, a normal transition can be made. At timestamp $2$, all the transitions going to the states $s_1$ and $s_2$ will be re-directed to the new states $s_1'$ and $s_2'$, denoting the states when the \textsc{Presence} happens. Other transitions that do not go  to the region will be performed normally. Similarly at time $3$, the transition from $s_3$ to $s_2$ will also go to the state $s_2'$ because the event is also true in this case. After time $4$, the original Markov transitions come back to work again. 
\end{example}

The intuition  can be formalized as follows. First, the original probabilities in $\mathbb{R}^{1\times m}$ is extended to $\mathbb{R}^{1\times 2m}$. Thus the initial probability $\boldsymbol\pi$ becomes $[\boldsymbol\pi,\textbf{0}]$.
Second, the transition matrix $\textbf{M}_{t}$ takes the form of four transition matrices between the two virtual worlds, i.e. the \textsc{Event} is true or false, in Equation (\ref{new-M-0}). Then the new transition matrix can be derived in Equations (\ref{new-M-1}) and (\ref{new-M-2})  for different time period  where $\textbf{M}$ is the original transition matrix and $\mathcal{\textbf{s}}^D$ is the diagonal of  the region $\mathcal{\textbf{s}}$ of \textsc{Presence} defined in Definition \ref{def-presence}.
\begin{myAlignS}
\label{new-M-0}
\textbf{M}_t=
\left[
\begin{array}{cc}
\textrm{false}\rightarrow \textrm{false}&\textrm{false}\rightarrow \textrm{true}\\
\textrm{true}\rightarrow \textrm{false}&\textrm{true}\rightarrow \textrm{true}\\
\end{array}
\right] \textrm{on the event.}
\end{myAlignS}
\begin{myAlignS}
\label{new-M-1}
\textbf{M}_t=
\left[
\begin{array}{cc}
\textbf{M}-\textbf{M}\textbf{s}^{\textbf{D}}&\textbf{M}\textbf{s}^{\textbf{D}}\\
\textbf{0}^{\textbf{D}}&\textbf{M}\\
\end{array}
\right], start-1\leq t\leq end-1.
\end{myAlignS}
\begin{myAlignS}
\label{new-M-2}
\textbf{M}_t=
\left[
\begin{array}{cc}
\textbf{M}&\textbf{0}^{\textbf{D}}\\
\textbf{0}^{\textbf{D}}&\textbf{M}\\
\end{array}
\right], t<start-1\ or\  t\geq end.
\end{myAlignS}
Equation (\ref{new-M-1}), designed to capture and maintain all the transitions going to the region of the $\textsc{Presence}$, is the new transition matrix when entering (and inside) the event time. Equation (\ref{new-M-2}), designed to keep the original transitions in the two virtual worlds, is the new transition matrix when leaving (and before) the event time. Third, at the last time $T$, the probability of the $\textsc{Presence}$ will be the sum of all probabilities in the bottom world (where $\textsc{Presence}$ is true).

\subsubsection{Pattern Events}
\begin{figure}[htbp]
	\centering
	\includegraphics[width=5cm]{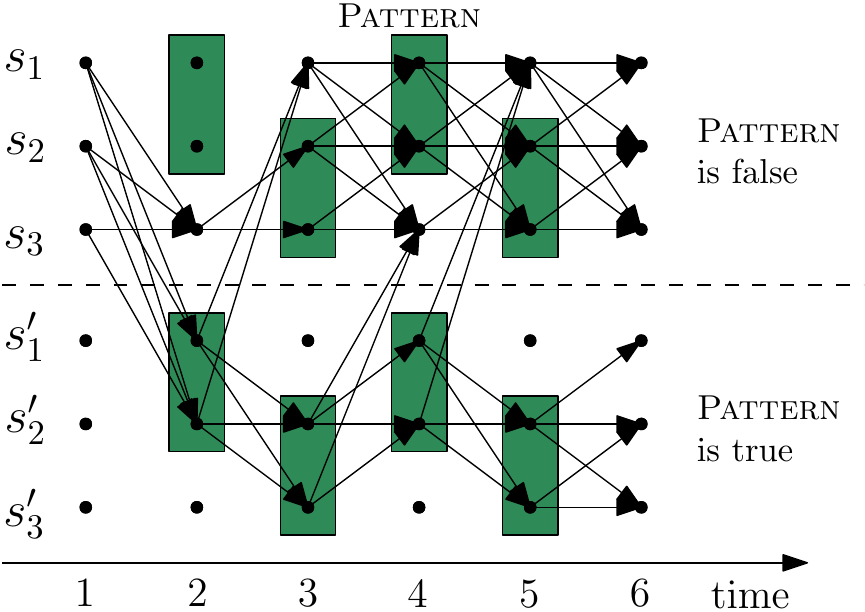}
	\caption{{\small New Markov transitions: at timestamp $1$, all transitions going to the defined region will be re-directed to the bottom world; at timestamp $2\sim 4$, only the transitions from the bottom world to the defined regions remain below.}}
	\label{Figure-example-prior-pattern}
\end{figure}

For \textsc{Pattern} events, the bottom world denoting the  event is true only needs to preserve the transitions going to the defined regions of the \textsc{Pattern} event. The following example shows the mechanism. 
\begin{example}
	\label{example-prior-pattern3}
	We study the \textsc{Pattern} event as illustrated in Fig.\ref{Figure-example-prior-pattern}. 
	At time $1$, the transitions entering $s_1$ and $s_2$ go to $s_1'$ and $s_2'$. From time $2$ to $4$, the transitions in the top world were performed  normally. However, the transitions from the bottom world go back to the top world if the destinations are not in  the defined regions. At time $5$, the  original Markov transitions come back to work again. 
\end{example}

From above example, the transition matrices for \textsc{Pattern}  differ from the ones for \textsc{Presence}  during the event time from $ start $ to $ end-1 $ (i.e., Equation (\ref{new-M-3})). 
On the other hand, when it is outside the event, i.e., $ t<start-1\ or\  t\geq end $, the transition matrices for \textsc{Pattern}  are the same as the ones for \textsc{Presence} (i.e., the matrices in \eqref{new-M-6} and \eqref{new-M-2} are the same).
Finally, when $ t=start-1 $, the transition matrices for \textsc{Pattern}  is also as same as the ones for \textsc{Presence} (i.e., the matrices in \eqref{new-M-4} and \eqref{new-M-1} are identical).
\begin{myAlignS}
\label{new-M-4}
\textbf{M}_t=
\left[
\begin{array}{cc}
\textbf{M}-\textbf{M}\textbf{s}^{\textbf{D}}&\textbf{M}\textbf{s}^{\textbf{D}}\\
\textbf{0}^{\textbf{D}}&\textbf{M}\\
\end{array}
\right], t=start-1.
\\
\label{new-M-3}
\textbf{M}_t=
\left[
\begin{array}{cc}
\textbf{M}&\textbf{0}^\textbf{D}\\
\textbf{M}-\textbf{M}\textbf{s}_t^{\textbf{D}}&\textbf{M}\textbf{s}_t^{\textbf{D}}\\
\end{array}
\right], start\leq t\leq end-1.
\\
\label{new-M-6}
\textbf{M}_t=
\left[
\begin{array}{cc}
\textbf{M}&\textbf{0}^{\textbf{D}}\\
\textbf{0}^{\textbf{D}}&\textbf{M}\\
\end{array}
\right], t<start-1\ or\  t\geq end.
\end{myAlignS}

In summary, the prior probability of any $\textsc{Event}$ can be derived as the sum of probabilities in the world where the $\textsc{Event}$ is true. Lemma \ref{theo-prior} shows the formal computation. 
\begin{lemma}
	\label{theo-prior}
	For initial probability $\boldsymbol\pi\in\mathbb{R}^{1\times m}$, the prior probability of an $\textsc{Event}$ of \textsc{Presence} and \textsc{Pattern}  is
	\begin{myAlignS}\Pr(\textsc{Event})=[\boldsymbol\pi , \textbf{0}] \prod_{i=1}^{end-1}\textbf{M}_i[\textbf{0},\textbf{1}]^\intercal\end{myAlignS}
	where $\textbf{M}_i$ is computed by Equations (\ref{new-M-1}), (\ref{new-M-2}), (\ref{new-M-4}), (\ref{new-M-3}), (\ref{new-M-6}).
\end{lemma}

If the Markov model is time-varying, i.e. when the transition matrices $ \textbf{M} $ at different $ t $ are not identical, the only extra effort is to re-compute Equations \eqref{new-M-1}$ \sim $\eqref{new-M-6} using the corresponding transition matrix $ \textbf{M} $ at $ t $.


\subsection{Computing Joint Probability of an Event}
\label{subsec:joint}
The calculation  of a spatiotemporal event and a sequence of observed locations, i.e, $ \Pr(o_1, o_2, \cdots, o _T, \textsc{EVENT}) $ is a little more complicated than previous sections since it depends on not only the initial probabilities but also the emission matrix of the LPPM.
Similarly, we use two-possible-world method to avoid enumerating all possible cases of an event. 
We utilize  forward-backward algorithm\cite{schusterbockler_introduction_2007} to estimate the probability of the true state (true location) at timestamp $t$ given all observations $\Pr(l_t|o_1,o_2,\cdots,o_T)$. 
It first calculates a forward probability $\alpha_t^k=Pr(l_t=s_k,o_1,o_2,\cdots,o_t)$ iteratively, i.e., 
\begin{myAlignS}
\alpha_t^k=Pr(o_t| l_t=s_k)\sum_{i}\alpha_{t-1}^i Pr(l_t=s_k|l_{t-1}=s_i).
\end{myAlignS}
Then, a backward probability {\footnotesize $\beta_t^k=$} {\footnotesize $\Pr(o_{t+1},o_{t+2},\cdots,o_T|l_t=s_k)$} can also be derived by
{ \begin{myAlignS}
{ \beta_t^k=\sum_i Pr(l_{t+1}=s_i|l_t=s_k)Pr(o_{t+1}|l_{t+1}=s_i)\beta_{t+1}^i}.
\end{myAlignS}}
By initializing $\beta_{T}^k=1$ for all $k$, we can obtain the estimation of $l_t$ as follows.
{ \begin{myAlignS}
{\footnotesize Pr(l_t=s_k|o_1,o_2,\cdots,o_T)=\frac{\alpha_t^k\beta_t^k}{\sum_i \alpha_t^i\beta_t^i}}
\end{myAlignS}}


\noindent{\bf Intuition.}
The intuition of our solution is to use the forward-backward algorithm in the two virtual worlds where the $\textsc{Event}$ is true and false. This is feasible because the emission probability, which determines the probabilities of the observations, is independent from any $\textsc{Events}$. Hence in our computation the forward probability and backward probability are $\Pr(\textsc{Event},o_1,o_2,\cdots,o_{t})$ for $t\leq end$ and $\Pr(o_{end+1},o_{end+2},\cdots,o_t|\textsc{Event})$ for $t>end$ respectively. 
By combining them together, we can obtain the posterior probability of the $\textsc{Event}$. Note that at any timestamp $t\leq end$, we do not see the future ($t>end$) observations. Thus the posterior probability only counts to the current timestamp $t$.

\noindent
\textbf{Before and During the Event.} 
In the forward algorithm, the probability 
$\alpha_t^k=\Pr(l_t=s_k,o_1,o_2,\cdots,o_t)$ is derived at timestamp $t$. We represent $\alpha_t^k$ in the vector form $\boldsymbol\alpha_t=[\alpha_1^1,\alpha_t^2,\cdots,\alpha_t^m]$. Then it can be derived as $\boldsymbol\alpha_t=(\boldsymbol\alpha_{t-1}\textbf{M}_{t-1})\circ\tilde{\textbf{p}}_{o_t}
=\boldsymbol\alpha_{t-1}\textbf{M}_{t-1}\tilde{\textbf{p}}_{o_t}^\textbf{D}$. Without any further observations, the joint probability can be derived from Lemma \ref{theo-prior}. The result is shown in Lemma \ref{lemma-post-before}.

\begin{lemma}
	\label{lemma-post-before}
	Given  initial probability $\boldsymbol\pi\in\mathbb{R}^{1\times m}$, the joint probability of an \textsc{Event} of $\textsc{Presence}$ or $\textsc{Pattern}$ and observations $o_1,o_2,\cdots,o_t$ at any timestamp $t\leq end$ is
	\begin{myAlignSSS}
	\label{eqn-post-before}
	\Pr(\textsc{Event},o_1,o_2,\cdots,o_t) 
	=[\boldsymbol\pi,\textbf{0}]\left(  \tilde{\textbf{p}}_{o_1}^\textbf{D}\prod_{i=2}^{t}(\textbf{M}_{i-1}\tilde{\textbf{p}}_{o_i}^\textbf{D})  \prod_{i=t}^{end-1}\textbf{M}_i   [\textbf{0},\textbf{1}]^\intercal \right)	\end{myAlignSSS}
\end{lemma}

\noindent
\textbf{After the Event.}
In the backward algorithm, {\footnotesize $\beta_t^k=\Pr(o_{t+1},o_{t+2},\cdots,o_T|l_t=s_k)$}. We represent it in the vector form {\footnotesize $\boldsymbol\beta_t=[\beta_t^1,\beta_t^2,\cdots,\beta_t^m]$}. 
Then it can be derived as {\footnotesize $\boldsymbol\beta_t=(\boldsymbol\beta_{t+1}\circ\tilde{\textbf{p}}_{o_{t+1}})\textbf{M}_t^\intercal 
=\boldsymbol\beta_{t+1}\tilde{\textbf{p}}_{o_{t+1}}^\textbf{D} \textbf{M}_t^\intercal$}  for any $t>end$. 
Similarly, we have Lemma \ref{lemma-post-after} for  joint probability.
\begin{lemma}
	\label{lemma-post-after}
	Given  initial probability $\boldsymbol\pi\in\mathbb{R}^{1\times m}$, the joint probability of an \textsc{Event} of $\textsc{Presence}$ or $\textsc{Pattern}$ and observations $o_1,o_2,\cdots,o_t$ at any timestamp $t>end$ is
	\begin{myAlignSSS}
	\label{eqn-post-after}
	&\Pr(\textsc{Event},o_1,o_2,\cdots,o_t)=[\boldsymbol\pi,\textbf{0}] \nonumber \\ 
	&\left(  \tilde{\textbf{p}}_{o_1}^\textbf{D}\prod_{i=2}^{end}(\textbf{M}_{i-1}\tilde{\textbf{p}}_{o_i}^\textbf{D})\right) \left( [\textbf{1},\textbf{1} ]\prod_{i=t-1}^{end}(\tilde{\textbf{p}}_{o_{i+1}}^\textbf{D}\textbf{M}_i^\intercal)\circ [\textbf{0},\textbf{1}]\right)^{\intercal}
	\end{myAlignSSS}
\end{lemma}
\vspace{-10pt}


To summarize, now we can quantify the  ratio
 {\small $\Pr(o_1,o_2,\cdots,o_T|\textsc{Event}) = \frac{Pr(o_1,o_2,\cdots,o_T, \textsc{Event}) }{Pr( \textsc{Event}) } $} for spatiotemporal event privacy using Lemma \ref{theo-prior} to compute {\small $ \Pr( \textsc{Event}) $} and Lemmas \ref{lemma-post-before}, \ref{lemma-post-after} to compute {\small $ \Pr(o_1,o_2,\cdots,o_T, \textsc{Event}) $}.
 We note that our approach of computing the joint probability of an event is able to deal with different emission matrices at each $ t $. Since {\footnotesize $\tilde{\textbf{p}}_{o_t}$} is a vector of emission probabilities given the observation $o_t$, i.e, a column in the emission matrix, and {\footnotesize $\tilde{\textbf{p}}_{o_t}^\textbf{D}$} is a diagonal matrix whose diagonal elements are  {\footnotesize $\tilde{\textbf{p}}_{o_t}$},  we  only need to obtain {\footnotesize $\tilde{\textbf{p}}_{o_t}$}  and {\footnotesize $\tilde{\textbf{p}}_{o_t}^\textbf{D}$} from the corresponding emission matrix at $ t $, and then use such {\footnotesize $\tilde{\textbf{p}}_{o_t}^\textbf{D}$} in Equations \eqref{eqn-post-before} and \eqref{eqn-post-after}.

\section{PriSTE framework}
\label{sec:priste}
In previous section, we designed methods for quantifying $ \epsilon $-spatiotemporal event privacy provided by an LPPM w.r.t. a specified  initial probability, which means that the privacy loss may not be bounded within $ \epsilon $  if an attacker has a different initial probability.

In this section, we first design the PriSTE (\underline{Pri}vate \underline{S}patio-\underline{T}emporal \underline{E}vent) framework and then solve the above problem by checking if   $ \epsilon $-spatiotemporal event privacy for any initial probabilities. 
Finally, we demonstrate two case studies that instantiate the framework based different location privacy metrics for protecting spatiotemporal event privacy.

\subsection{PriSTE}

\begin{figure}[h]
	\centering
	\includegraphics[width=8.5cm]{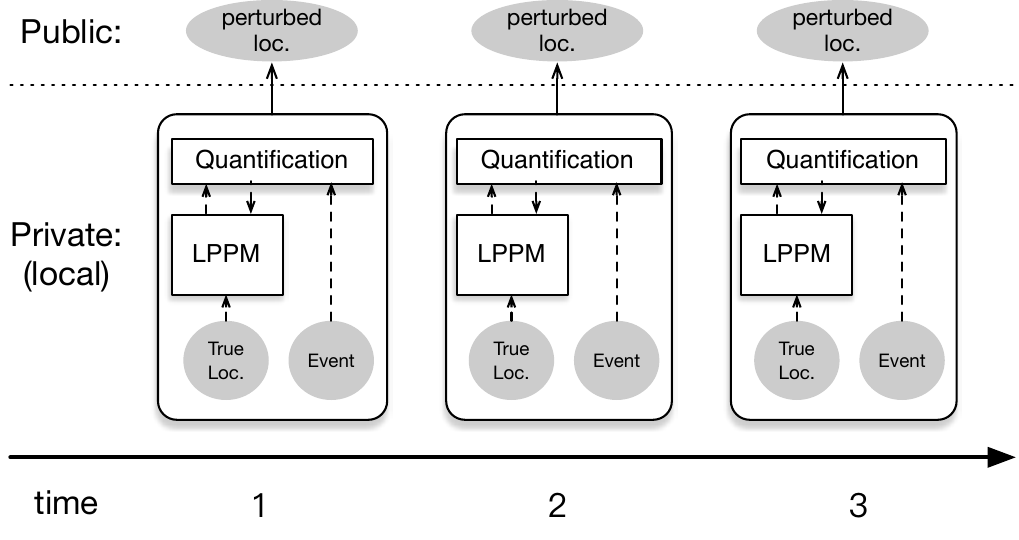}
	\caption{{\small PriSTE framework.}}
	\label{fig:fm}
\end{figure}

Based on the quantification techniques that we developed in previous sections, we propose a framework that converts a location privacy protection mechanism  into one protecting spatiotemporal event privacy.
The PriSTE framework is illustrated in Fig.\ref{fig:fm} and described in Algorithm \ref{alg-framework}.

The major components are \textit{Quantification} and a given LPPM.
Their interactions are described as follows.
First, the LPPM generates a perturbed location from the true location (Line 2 in Algorithm \ref{alg-framework})  and pass it to the \textit{Quantification} component.
By Theorem \ref{theo-eps-delta-DP}, the \textit{Quantification} component (Line 3)  checks whether this perturbed location satisfies the ratio in Equation \eqref{eqn-eps-delta-DP1} (i.e., $ \epsilon $-spatiotemporal event privacy),  under a sequence of previous observations and user-specified spatiotemporal events.
If not, we need to calibrate the emission matrix to ensure that it satisfies $ \epsilon $-spatiotemporal event privacy. 
The strategy of emission matrix calibration is  LPPM-dependent.
In the next section, we demonstrate case studies of \geoind\cite{andres_geo-indistinguishability:_2013} and {$\delta$-location set privacy}\cite{xiao_protecting_2015}, which are the state-of-the-art location  privacy metrics.


\begin{algorithm}[h]
	\caption{PriSTE Framework}
	\begin{algorithmic}[1]
		\Require{ true location,
			$\epsilon$, LPPM, $\textbf{M}$, \textsc{Events}
		}
		\For{$t$ in $\{1,2,\cdots,T\}$}
		\State{generate $o_t$  with LPPM w.r.t. the true location;}
		\While{$\epsilon$-Spatiotemporal Event Privacy not hold}
		\label{alg-if-privacy}
		\State{\textit{calibrate} LPPM and generate $o_t$;}		
		\EndWhile
		\State{release $o_t$;}
		\EndFor
	\end{algorithmic}
	\label{alg-framework}
\end{algorithm}

\subsection{Privacy Checking with Arbitrary Initial Probability}
\label{subsec:privacycheck}

According to the quantification in Section \ref{sec: quantifying}, we can calculate {\small $\frac{\Pr(o_1,o_2,\cdots,o_T|\textsc{Event})}{
		\Pr(o_1,o_2,\cdots,o_T|\lnot\textsc{Event})}$} given  $ o_1,o_2, \cdots,o_T $ and a given initial probability {$ {\boldsymbol\pi} $}.
In this section, we show how to make sure the ratio is bounded given arbitrary initial probability.

Our idea to  is taking $ {\boldsymbol\pi} $ as a variable and solving the maximization problem of {\small $\frac{\Pr(o_1,o_2,\cdots,o_T|\textsc{Event})}{\Pr(o_1,o_2,\cdots,o_T|\lnot\textsc{Event})}- e^\epsilon $}.
We want to make sure the maximum value is always less than or equal to $0$, i.e., the user enjoys plausible deniability for her specified spatiotemporal event.

%
%

The following theorem shows the conditions related to $ {\boldsymbol\pi} $ that satisfies $ \epsilon $-spatiotemporal event privacy.  
We will  formulate it as an optimization problem. 

\begin{theorem}
	\label{theo-eps-delta-DP}
	For an $\textsc{Event}$ of \textsc{Presence} or \textsc{Pattern} and an arbitrary initial probability $\boldsymbol\pi\in\mathbb{R}^{1\times m}$, 
	 $\epsilon$-spatiotemporal event privacy  is satisfied at any timestamp $t$, i.e., {\small $\frac{\Pr(o_1,o_2,\cdots,o_T|\textsc{Event})}{\Pr(o_1,o_2,\cdots,o_T|\lnot\textsc{Event})} \leq e^\epsilon$},
	if the observation $o_t$ is released based on the following two conditions
	\begin{myAlignSSS}
	\label{eqn-DP-check1}
	\boldsymbol\pi \left( [\textbf{1}^\textbf{D},\textbf{0}^{\textbf{D}}] \left( (e^\epsilon-1)\textbf{a}^{\intercal}\textbf{b}-e^{\epsilon}\textbf{a}^{\intercal}\textbf{c} \right)[\textbf{1}^\textbf{D},\textbf{0}^{\textbf{D}}]^{\intercal} \right) \boldsymbol\pi^\intercal 
	+\boldsymbol\pi[\textbf{1}^\textbf{D},\textbf{0}^{\textbf{D}}]{\left( \textbf{b}^\intercal \right)}\leq 0
	\end{myAlignSSS}
	\begin{myAlignSSS}
	\label{eqn-DP-check2}
	\boldsymbol\pi \left([\textbf{1}^\textbf{D},\textbf{0}^{\textbf{D}}] \left( (e^\epsilon-1)\textbf{a}^{\intercal}\textbf{b}+\textbf{a}^{\intercal}\textbf{c} \right)[\textbf{1}^\textbf{D},\textbf{0}^{\textbf{D}}]^{\intercal} \right)\boldsymbol\pi^\intercal
	-\boldsymbol\pi[\textbf{1}^\textbf{D},\textbf{0}^{\textbf{D}}]{\left( e^{\epsilon}\textbf{b}^\intercal\right)}\leq 0
	\end{myAlignSSS}
	 where
	\begin{myAlignSS}	\textbf{a}^\intercal=\prod_{i=1}^{end-1}\textbf{M}_i[\textbf{0},\textbf{1}]^\intercal
	\end{myAlignSS}
	For $t\leq end$,
	\begin{myAlignSS}
	\textbf{b}^\intercal= \tilde{\textbf{p}}_{o_1}^\textbf{D}\prod_{i=2}^{t}(\textbf{M}_{i-1}\tilde{\textbf{p}}_{o_i}^\textbf{D})  \prod_{i=t}^{end-1}\textbf{M}_i   [\textbf{0},\textbf{1}]^\intercal  \\
	\textbf{c}^\intercal=  \tilde{\textbf{p}}_{o_1}^\textbf{D}\prod_{i=2}^{t}(\textbf{M}_{i-1}\tilde{\textbf{p}}_{o_i}^\textbf{D})    [\textbf{1},\textbf{1}]^\intercal 
	\end{myAlignSS}
	For $t>end$,
	\begin{myAlignSS}
	\textbf{b}^\intercal=\tilde{\textbf{p}}_{o_1}^\textbf{D}\prod_{i=2}^{end}(\textbf{M}_{i-1}\tilde{\textbf{p}}_{o_i}^\textbf{D}) \left([\textbf{1},\textbf{1} ]\prod_{i=t-1}^{end}(\tilde{\textbf{p}}_{o_{i+1}}^\textbf{D}\textbf{M}_i^\intercal) \circ [\textbf{0},\textbf{1}]\right)^\intercal
	\\
	\textbf{c}^\intercal=\tilde{\textbf{p}}_{o_1}^\textbf{D}\prod_{i=2}^{end}(\textbf{M}_{i-1}\tilde{\textbf{p}}_{o_i}^\textbf{D}) \left([\textbf{1},\textbf{1} ]\prod_{i=t-1}^{end}(\tilde{\textbf{p}}_{o_{i+1}}^\textbf{D}\textbf{M}_i^\intercal) \circ [\textbf{1},\textbf{1}]\right)^\intercal 
	\end{myAlignSS}
\end{theorem}

\noindent{\bf Quadratic Programming.} 
To determine whether Equations (\ref{eqn-DP-check1}) and (\ref{eqn-DP-check2}) are true or not for arbitrary  $ \boldsymbol{\pi} $, we transform them to maximization problems: finding the maximum values of the left parts of Equations (\ref{eqn-DP-check1}) and (\ref{eqn-DP-check2}) under the constraints of $ 0 \leq p_i  \leq 1$ where $ p_i \in \boldsymbol{\pi}$.
As long as one maximum value is larger than $ 0 $,  we know that the LPPM  (emission matrix) may not satisfy $ \epsilon $-spatiotemporal event privacy.
The maximization are equivalent to quadratic programing problem since they can be rewritten in a form of {\small $\boldsymbol\pi\textbf{A}\boldsymbol\pi^{\intercal}=\frac{1}{2}\boldsymbol\pi(\textbf{A}+\textbf{A}^{\intercal})\boldsymbol\pi^{\intercal}$} where  {\small \textbf{A} } is a matrix. 
 We skip  the computation details about solving such quadratic programing problem since many methods and tools  have been  proposed in  literature.
 In the experiments, we use IBM CPLEX optimizer \cite{cplex} as our computation engine. 

\subsection{Case Study 1: PriSTE with Geo-indistinguishability}
\label{sec:case_study1}
In this section, we  instantiate PriSTE framework using $\alpha$-Planar Laplace mechanism ($\alpha$-PLM) which is designed for Geo-indistinguishability\cite{andres_geo-indistinguishability:_2013}.
We first show the computation details for quantifying $ \epsilon $-spatiotemporal event privacy by Theorem \ref{theo-eps-delta-DP}, and then design a greedy strategy for approximately achieving $ \epsilon $-spatiotemporal event privacy.
  
{\bf Algorithm Design.} 
To implement  the quantification component, we need to  (1) compute the internal parameters $\textbf{a}$, $\textbf{b}$ and $\textbf{c}$  shown in Theorem \ref{theo-eps-delta-DP}, and (2)  design a strategy to calibrate the emission matrix.

For the calibration strategy for Planar Laplace Mechanism (PLM) with a specified privacy budget $ \alpha $ (which solely determines the shape of the output distribution), we  exponentially decay its privacy budget  because a smaller privacy budget implies stronger protection for location privacy and less information disclosure.
In our algorithm, decay rate $\frac{1}{2}$ for the privacy budget in line \ref{alg-budget-halve} of Algorithm \ref{alg-priv-check1} is a tunable parameter that provides a trade-off between efficiency and utility of the released locations.  
Setting a small value allows the algorithm converge faster, but at the cost of over-perturbing the location at each timestamp. In contrast, using a large value is less efficient but allows better utility to be achieved.

A natural question is whether we can  always find an $\alpha$ to release a perturbed location that satisfies Equation \eqref{eqn-eps-delta-DP1}. 
The answer is affirmative because $\alpha$ converges exponentially to $0$. When $\alpha=0$, it releases no useful information about the true location, i.e., uniformly returning a random location without using user's true position.
It is easy to verify that the Equations \eqref{eqn-DP-check1} and \eqref{eqn-DP-check2} are always true in this situation. 

Algorithm \ref{alg-priv-check1} shows the computation process. 
To boost the efficiency of our algorithm, we use intermediate matrices $\textbf{A}$ and $\textbf{B}$ to facilitate the computation of $\textbf{b}$ and $\textbf{c}$. 
At time $1$, we initialize the variables as line $\ref{alg-line-t1-start}\sim \ref{alg-line-t1-end}$. 
At any time before and inside the $\textsc{Event}$, we compute the variables as line $\ref{alg-line-in-start} \sim \ref{alg-line-in-end}$. 
At any timestamps after the $\textsc{Event}$, the variables are derived as line $\ref{alg-line-out-start}\sim \ref{alg-line-out-end}$. 
Then we use quadratic programming methods to check Eq.(\ref{eqn-DP-check1}) and (\ref{eqn-DP-check2}) to decide whether to release the $o_{t}$ or not. 
If not, we generate a new $o_{t}$ with only half $\alpha$, and repeat the above process again.
Finally, we update the matrices $\textbf{A}$ and $\textbf{B}$ as line {\small $\ref{alg-line-AB-start}\sim\ref{alg-line-AB-end}$}. 
If $t=end$,  in line \ref{alg-line-in-start}, the product {\footnotesize $\prod_{i=t}^{end-1}\textbf{M}_i$} will be the identity matrix. 
In line \ref{alg-line-update-A}, {\footnotesize $\textbf{M}_{0}$} is the identity matrix when $t=1$.
We note that for multiple \textsc{Events}, Algorithm \ref{alg-priv-check1} can be executed multiple times for each \textsc{Event}. 
\begin{algorithm}[h]
	\caption{PriSTE with Geo-indistinguishability.}
	\begin{algorithmic}[1]
		\Require{
			$\epsilon$, \textsc{Event}, $\alpha$-PLM, $\textbf{M}_i$, $\forall i\in\{1,2,\cdots,T\}$
		}
		\For{$t$ in $\{1,2,\cdots,T\}$}
		\State{$o_t\gets$ $\alpha$-PLM;}\Comment{{\tt \scriptsize initial budget$=\alpha$}}
		\label{alg-line-ot}
		\If{$t==1$}
		\State{$\textbf{a}^\intercal\gets  \prod_{i=1}^{end-1}\textbf{M}_i[\textbf{0},\textbf{1}]^\intercal$}
		\label{alg-line-t1-start}
		\State{$\textbf{A}\gets\textbf{I} $}
		\Comment{{\tt \scriptsize identity matrix}}
		\State{$\textbf{B}\gets\textbf{I}$}
		\State{$\textbf{b}^\intercal\gets   \tilde{\textbf{p}}_{o_1}^\textbf{D} \textbf{a}^\intercal$}
		\State{$\textbf{c}^\intercal\gets \tilde{\textbf{p}}_{o_1}^{\intercal} $}
		\label{alg-line-t1-end}
		\ElsIf{$t<=end$}
		\Comment{{\tt \scriptsize before and during \textsc{Event}}}
		\State{$\textbf{b}^\intercal\gets \textbf{A}\textbf{M}_{t-1} \tilde{\textbf{p}}_{o_t}^\textbf{D} \prod_{i=t}^{end-1}\textbf{M}_i [\textbf{0},\textbf{1}]^\intercal$}
		\label{alg-line-in-start}
		\State{$\textbf{c}^\intercal\gets \textbf{A}\textbf{M}_{t-1}\tilde{\textbf{p}}_{o_t}^\textbf{D}  [\textbf{1},\textbf{1}]^\intercal$}
		\label{alg-line-in-end}
		\Else
		\Comment{{\tt \scriptsize after \textsc{Event}}}
		\State{$\textbf{b}^\intercal\gets \textbf{A}\left(([\textbf{1},\textbf{1} ]\tilde{\textbf{p}}_{o_t}^\textbf{D}\textbf{M}_{t-1}^\intercal \textbf{B})\circ[\textbf{0},\textbf{1}]\right)^\intercal$}
		\label{alg-line-out-start}
		\State{$\textbf{c}^\intercal\gets \textbf{A}\left(([\textbf{1},\textbf{1} ]\tilde{\textbf{p}}_{o_t}^\textbf{D}\textbf{M}_{t-1}^\intercal \textbf{B})\circ[\textbf{1},\textbf{1}]\right)^\intercal$}
		\label{alg-line-out-end}
		\EndIf
		\If{Equations (\ref{eqn-DP-check1}) and (\ref{eqn-DP-check2}) hold} \Comment{{\tt \scriptsize $\epsilon$ is used here.}}
		\label{alg-line-if-check}
		\State{release $o_t$;}
		\Comment{{\tt \scriptsize okay to  release $o_t$}}
		\Else
		\State{$\alpha\gets \frac{\alpha}{2}$, goto Line \ref{alg-line-ot};}
		\label{alg-budget-halve}
		\Comment{{\tt \scriptsize halve the budget}}
		\EndIf
		\If{ $t\leq end$}
		\label{alg-line-AB-start}
		\State{$\textbf{A}\gets \textbf{A}\textbf{M}_{t-1}\tilde{\textbf{p}}_{o_t}^\textbf{D}$}
		\label{alg-line-update-A}
		\Comment{{\tt \scriptsize update $\textbf{A}$  by the real $o_{t}$}}
		\Else
		\State{$\textbf{B}\gets\tilde{\textbf{p}}_{o_t}^\textbf{D} \textbf{M}_{t-1}^\intercal  \textbf{B}$}
		\Comment{{\tt \scriptsize update $\textbf{B}$  by the real $o_{t}$}}
		\EndIf
		\label{alg-line-AB-end}
		\EndFor
		
		%
	\end{algorithmic}
	\label{alg-priv-check1}
\end{algorithm}

%
%

\noindent{\bf Complexity.} 
The internal parameters $\textbf{a}$, $\textbf{b}$ and $\textbf{c}$ in Algorithm \ref{alg-priv-check1} need $O(mT)$ time to be evaluated. 
The major computational cost lies in the quadratic program for checking Equations (\ref{eqn-DP-check1}) and (\ref{eqn-DP-check2}). 
The complexity will be determined by the quadratic matrix {\footnotesize $[\textbf{1}^\textbf{D},\textbf{0}^{\textbf{D}}]\textbf{a}^\intercal\textbf{c}[\textbf{1}^\textbf{D},\textbf{0}^{\textbf{D}}]^{\intercal}$}. 
If it is positive definite, then the complexity is $O(m^{3})$. 
Otherwise, with any negative eigenvalues, it will be NP-hard \cite{pardalos_quadratic_1991}. 
In our experiments, we use IBM CPLEX  which can provide globally optimal results for quadratic program but may need a long computation time.
We  use a \textit{conservative release} strategy to remedy this:
we use a threshold to limit the computation time of quadratic program for checking Eq.\eqref{eqn-DP-check1} and \eqref{eqn-DP-check2}.
It will not release a perturbed location unless the equations are true.
Although it may lead to suboptimal solution in budget calibration, it always guarantees $\epsilon$-spatiotemporal event privacy since every released locations satisfy Eq.\eqref{eqn-DP-check1} and \eqref{eqn-DP-check2}.

\noindent{\bf Privacy Analysis.}
PriSTE framework relies on a local model, i.e., the assumption that adversaries cannot obtain user's locally stored  information as shown in Fig.\ref{fig:fm}.
Although line \ref{alg-line-ot} may be executed more than once at a timestamp $ t $, Algorithm \ref{alg-priv-check1} still satisfies $ \alpha ' $-geo-indistinguishability where $ \alpha ' $ is the final privacy budget used for releasing $ o_t $ because that is the only observation of attacker at time $ t $.
If we remove the assumption of local model, the above statements may not be true since attacker may observe the internal states of the algorithm (which is the privacy goal of \textit{pan-privacy}\cite{dwork_pan-private_2010}).
Examples of  internal states includes  multiple  $ o_t $ tested at $ t $ or the final $ \alpha ' $ used in the algorithm.
Anothe assumption that may affect the privacy guarantee is the transition matrix \textbf{M}, which we use it to model the correlations between locations and assume that it is given.
It is  an interesting future work to quantify the change of privacy loss in terms of $ \epsilon $-spatiotemporal event privacy  if the ground truth of correlation is not the modeled one.
We defer this study to future work.
 

\subsection{Case Study 2: PriSTE with $\delta$-Location Set Privacy}
\label{sec:case_study2}
To evaluate the effectiveness of PriSTE under different location privacy protection mechanisms, we also instantiate it using another  privacy metric \textit{$\delta$-location set privacy}\cite{xiao_protecting_2015}\cite{xiao_loclok:_2017}, which is proposed for obtaining better utility by taking advantage of temporal correlation between consecutive locations in user's trajectory.
The key idea is that hiding the true location in any impossible locations (e.g., whose probabilities are close to 0) is a lost cause because the adversary already knows the user cannot be there.
In other words, it restricts the output domain of the emission matrix to \textit{$\delta$-location set}, which is a set containing minimum number of locations that have prior probability sum no less than $1-\delta$.
A larger $\delta$ indicates weaker privacy guarantee.

The privacy metrics of  $ \alpha $-geo-indistinguishability and $\delta$-location set privacy  are orthogonal because the former requires a specific ``shape'' of emission distribution and the latter restricts output domain of the emission distribution.
In \cite{xiao_protecting_2015}, Xiao and Xiong proposed a framework to achieve $\delta$-location set privacy using a given LPPM.
For ease of comparison, we use $ \alpha $-PLM as the underlying LPPM for $\delta$-location set privacy.

\begin{algorithm}[h]
	\caption{PriSTE with $\delta$-Location Set Privacy.}
	\begin{algorithmic}[1]
		\Require{
			$\epsilon$, \textsc{Event}, $\alpha$-PLM, $\textbf{M}_i$, $\forall i\in\{1,2,\cdots,T\}$, $\pi$, $\delta$, $ \mathbf{M}$.
		}
		\For{$t$ in $\{1,2,\cdots,T\}$}
		\State{$ \mathbf{p}_t^-  \gets \mathbf{p}_{t-1}^+ \mathbf{M} $;}\Comment{{\tt \scriptsize Markov transition}}
		\State{Construct $ \Delta \mathbf{X}_t $}\Comment{{\tt \scriptsize $\delta$-location set}}
		\State{$o_t\gets$ $\alpha$-PLM within $ \Delta \mathbf{X}_t $;}
		\State{the same as Lines 3 $ \sim $ 15 in Algorithm \ref{alg-priv-check1};}
		\If{Equations (\ref{eqn-DP-check1}) and (\ref{eqn-DP-check2}) hold} \Comment{{\tt \scriptsize $\epsilon$ is used here.}}
		\label{alg-line-if-check}
		\State{release $o_t$;}\Comment{{\tt \scriptsize okay to  release $o_t$}}
		\State{Derive posterior probability $ \mathbf{p}_t^+ $ by Eq.\eqref{eq:posterior};}
		\Else
		\State{$\alpha\gets \frac{\alpha}{2}$, goto Line 4;}
		\label{alg-budget-halve}
		\Comment{{\tt \scriptsize halve the budget}}
		\EndIf
		\State{the same as Lines 21 $ \sim $ 25 in Algorithm \ref{alg-priv-check1};}
		\EndFor
	\end{algorithmic}
	\label{alg-priv-check2}
\end{algorithm}

In Line 2, when $ t=1 $, we have {\footnotesize $ \mathbf{p}_0^+ = \pi $}.
In Line 8, according to \cite{xiao_protecting_2015}, the posterior probability can be calculated by  Equation \eqref{eq:posterior} where {\footnotesize $ \mathbf{p}_t^+[j] $} and {\footnotesize $ \mathbf{p}_t^-[i] $} are the $ i $th elements in the corresponding probability vectors.
\begin{myAlignSSS}
	\mathbf{p}_t^+[i] = \Pr(l_t = s_i|o_t) = \frac{\Pr(o_t|l_t = s_i)*\mathbf{p}_t^-[i]}{\sum_{j} \Pr(o_t|l_t = s_j)*\mathbf{p}_t^-[j]} \label{eq:posterior}
\end{myAlignSSS}

 Hence, we need the initial probability $\pi$ in order to calculate $\delta$-location set. 
In experiments, we set $\pi$ to a uniform distribution for the evaluation of $\delta$-location set privacy.
We note that  PriSTE  is  agnostic to such initial probability since it guarantees spatiotemporal event privacy against adversaries with arbitrary  knowledge about the initial probability.

\section{Experimental Evaluation}
\label{sec:exp}
In experiments, we verified that Algorithms \ref{alg-priv-check1}  and  \ref{alg-priv-check2}  can adaptively calibrate the privacy budget of Planar Laplace Mechanism (PLM) at each timestamp for  both location privacy and spatiotemporal event privacy.
Especially, we highlight the following empirical findings.
\begin{itemize}
\item A stricter LPPM can satisfy a certain level of spatiotemporal event privacy \textit{without} any change (i.e., no need of privacy budget calibration), whereas a more loose LPPM may need to reduce its privacy budget significantly for protecting the same event.
\item For achieving the same level of $\epsilon$-spatiotemporal event privacy using different LPPMs, a stricter LPPM is \textit{not} always better in terms of data utility.
\item If the user's transition matrix has a significant pattern, an LPPM may need a small privacy budge to achieve $ \epsilon $-spatiotemporal event privacy.
\end{itemize}

\subsection{Experiment Settings and Metrics}
\noindent{\bf Dataset.} We used real-life and synthetic datasets in experiments.
	Geolife data \cite{zheng_geolife:_2010} was collected from $182$ users in a period of over three years.
	It recorded a wide range of users' outdoor movements, represented by a series of tuples containing latitude, longitude and timestamp.
	The user's entire trajectory is used to train the transition matrix $\textbf{M}$, e.g. with R package ``markovchain''. 
	
	We generated a synthetic trajectory and its transition probability matrix as follows. First, a map with $20*20$ cells is generated. Then, the transition probability from one cell to another is drawn from the two-dimensional Gaussian distribution  with scale parameter $\sigma$ based on the distance between the cells.
	Here, a smaller   $\sigma$ indicates that the user moves to the adjacent cells with a higher probability, i.e., the transition matrix has a more significant pattern.
	Finally, we produced trajectories with $50$ timestamps using such transition matrix to simulate  movement of a user.

%
%

\noindent{\bf Quadratic Programming.} We use the IBM CPLEX optimizer \cite{cplex} (version 12.7.1) to find the globally optimal solution for the quadratic programming in Algorithm \ref{alg-priv-check1}. 
We adopt a  strategy of \textit{conservative release} as mentioned previously and limit  the computation time for each optimization to 1 second.

\noindent{\bf \textsc{Events}. }
We investigate  $\textsc{Presence}$ and $\textsc{Pattern}$ events, which are represented by two parameters $\mathcal{S}$ and $\mathcal{T}$. For example, {\footnotesize $\textsc{Presence}({\mathcal{S}}=\{1:10\},\mathcal{T}=[4:8])$} is \textsc{Presence} event denoting the user appears in the region of {\footnotesize $\{s_{1},s_{2},\cdots,s_{10}\}$} during timestamps {\footnotesize $\{4,5,6,7,8\}$}.

\noindent{\bf Utility Metrics.} 
We use two metrics to evaluate data utility.
\begin{itemize}
	\item
	Privacy budget $\alpha$ used in PLM, including  $\alpha$ at each timestamp (see Section \ref{sec:utility1}) and the average $\alpha$ during the whole time period (see Section \ref{sec:utility2}).
	A higher privacy budget indicates  higher utility.
	\item
	The Euclidean distance between the perturbed locations and the true locations.
	A smaller Euclidean distance indicates  higher utility. 
\end{itemize}
We run our algorithm  $100$ times and aggregate the results to calculate average privacy budget and  Euclidean distance.

\subsection{Utility at Each Timestamp}
\label{sec:utility1}
In this section, we show the utility (average privacy budget over 100 runs) at each timestamp  for protecting {\footnotesize $\textsc{Presence}(\mathcal{S}=[1:10],\mathcal{T}=[4:8])$} and {\footnotesize $\textsc{Presence}(\mathcal{S}=[1:10],\mathcal{T}=[16:20])$}.

\begin{figure}[h]
	\centering
	\begin{subfigure}{0.24\textwidth}
		\centering
		\includegraphics[width=4.0cm]{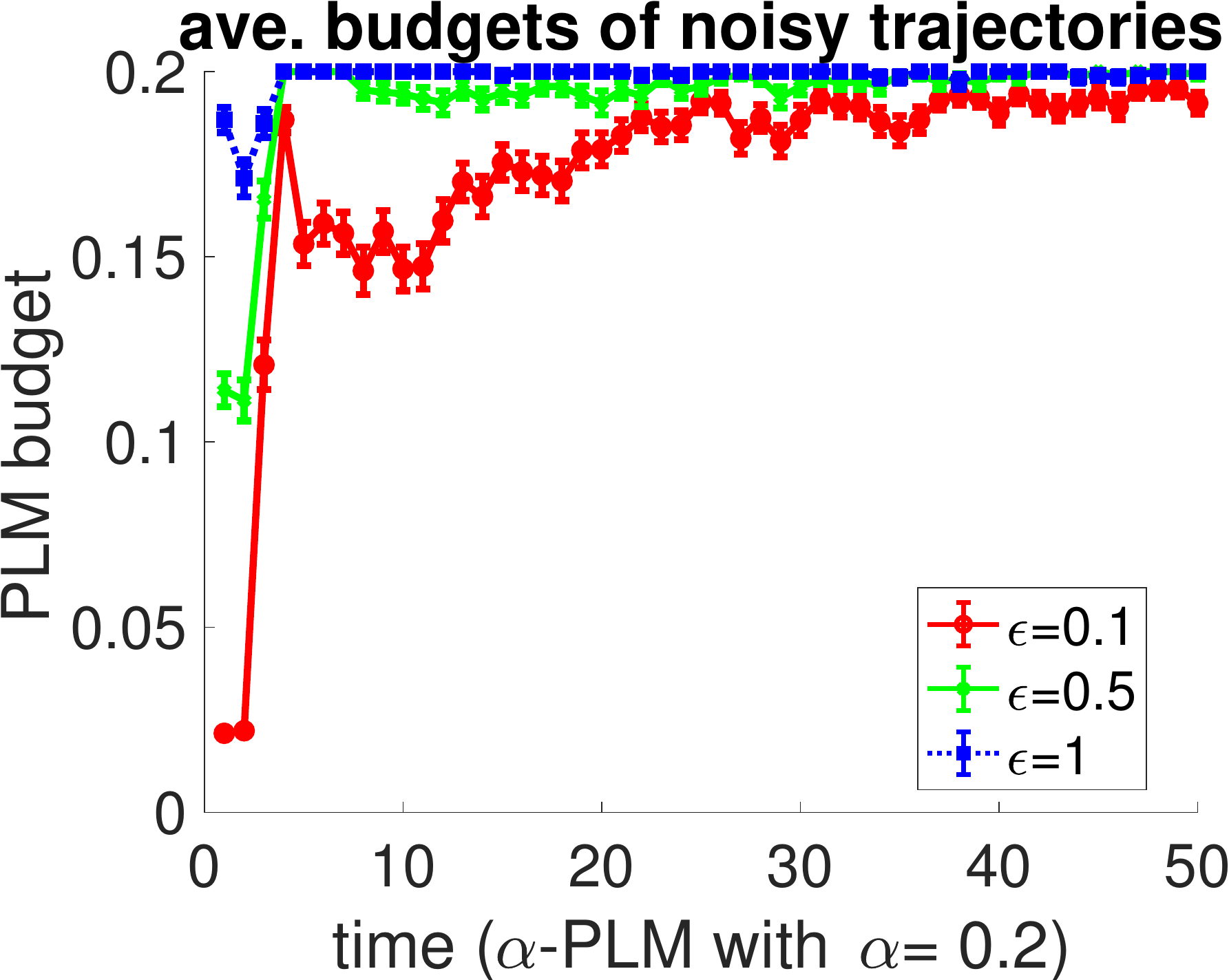}
		\vspace{-6pt}
		\caption{{\scriptsize $ 0.2 $-PLM  for different  $\epsilon$.}}
		\label{Figure-expmt-B11}
	\end{subfigure}
	\begin{subfigure}{0.24\textwidth}
		\centering
		\includegraphics[width=4.0cm]{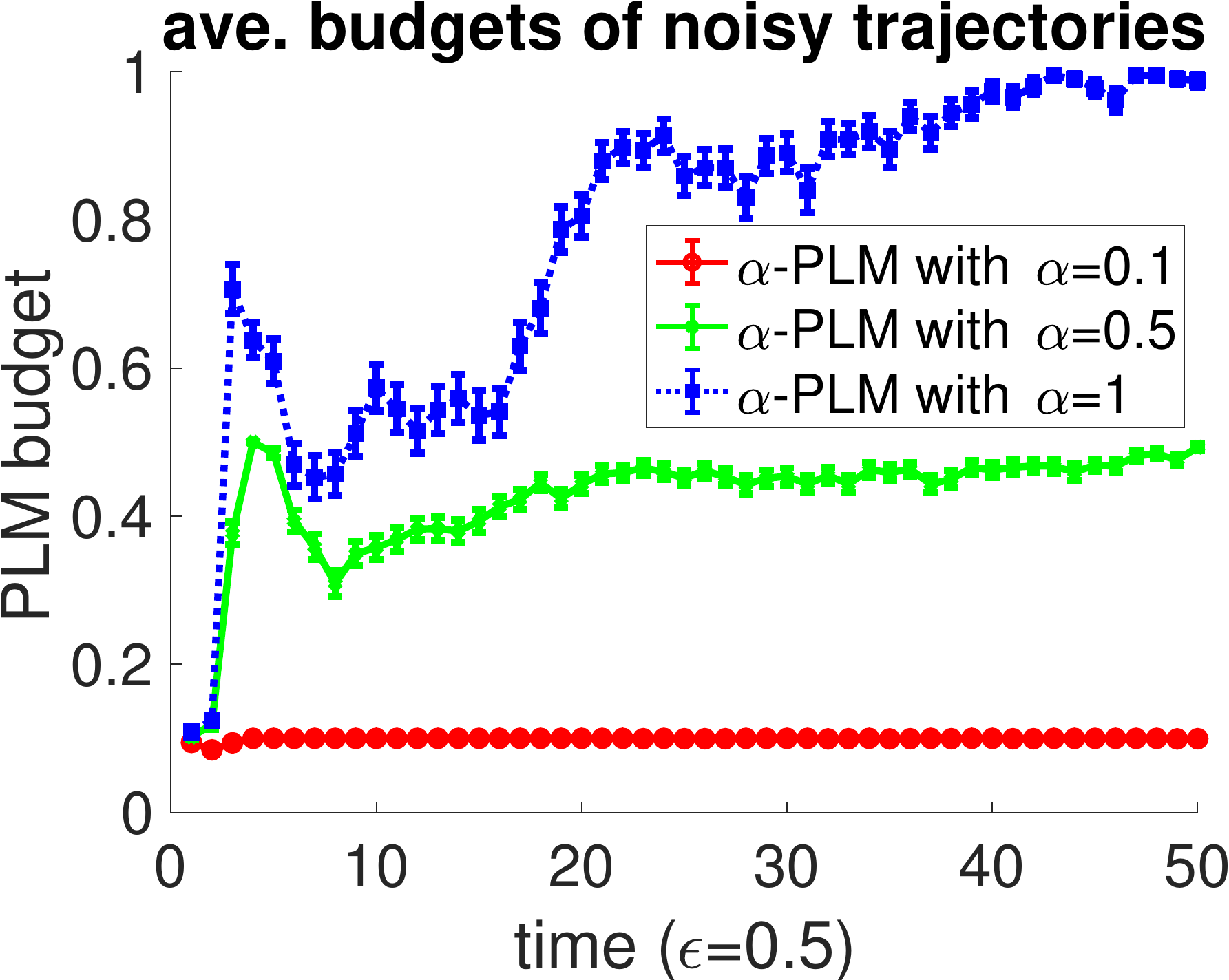}
			\vspace{-6pt}
		\caption{{\scriptsize  Different PLMs for $\epsilon=0.5$.}}
		\label{Figure-expmt-B22}
	\end{subfigure}
	\caption{{\footnotesize  $\textsc{Presence}(\mathcal{S}=[1:10],\mathcal{T}=[4:8])$ on synthetic data.}}
	\label{Figure-expmt-B11-22}
\end{figure}

\begin{figure}[!htbp]
	\centering
	\begin{subfigure}{0.24\textwidth}
		\centering
		\includegraphics[width=4.0cm]{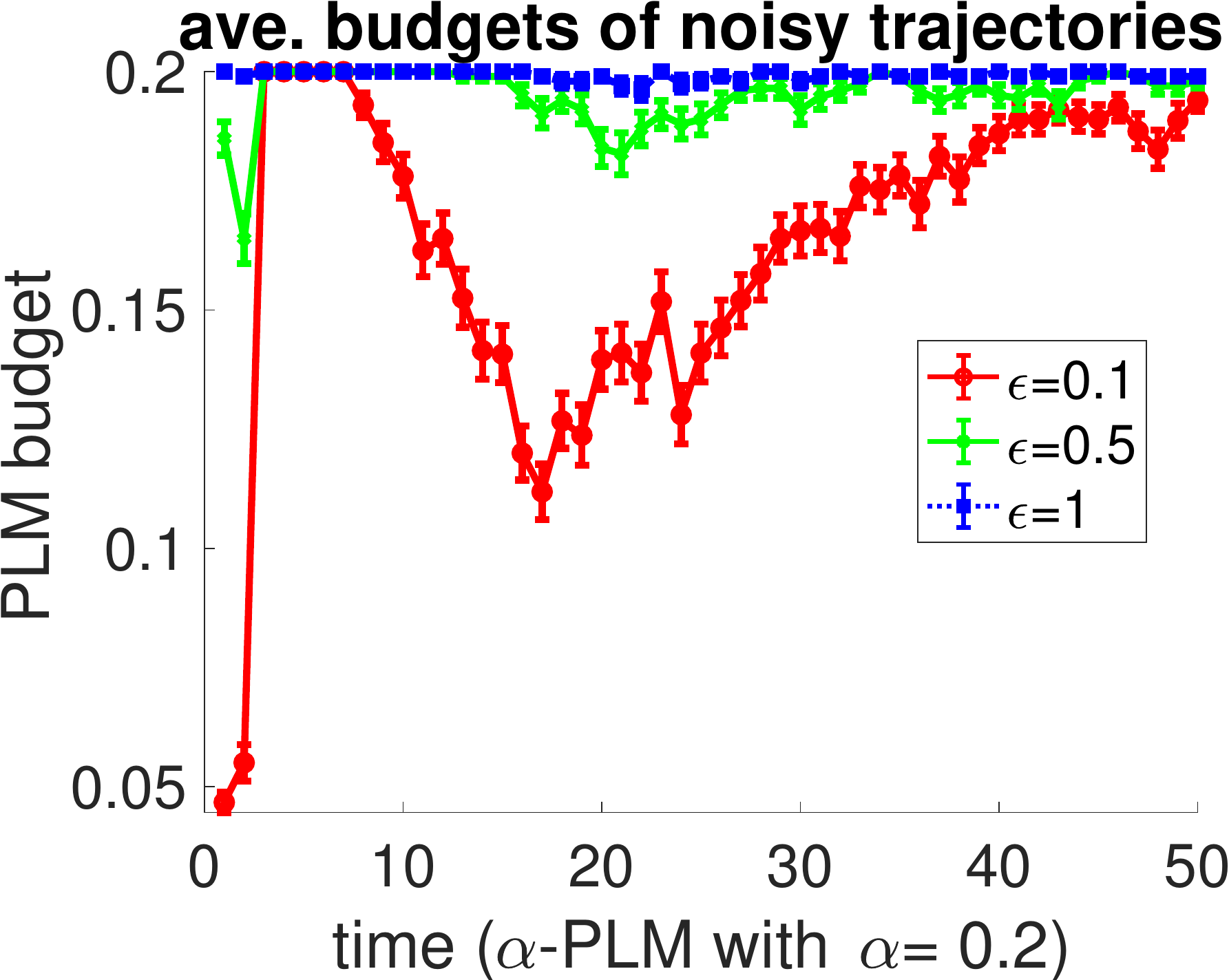}
			\vspace{-6pt}
		\caption{{\scriptsize $ 0.2 $-PLM  for different  $\epsilon$.}}
		\label{Figure-expmt-B33}
	\end{subfigure}
\begin{subfigure}{0.24\textwidth}
	\centering
	\includegraphics[width=4.0cm]{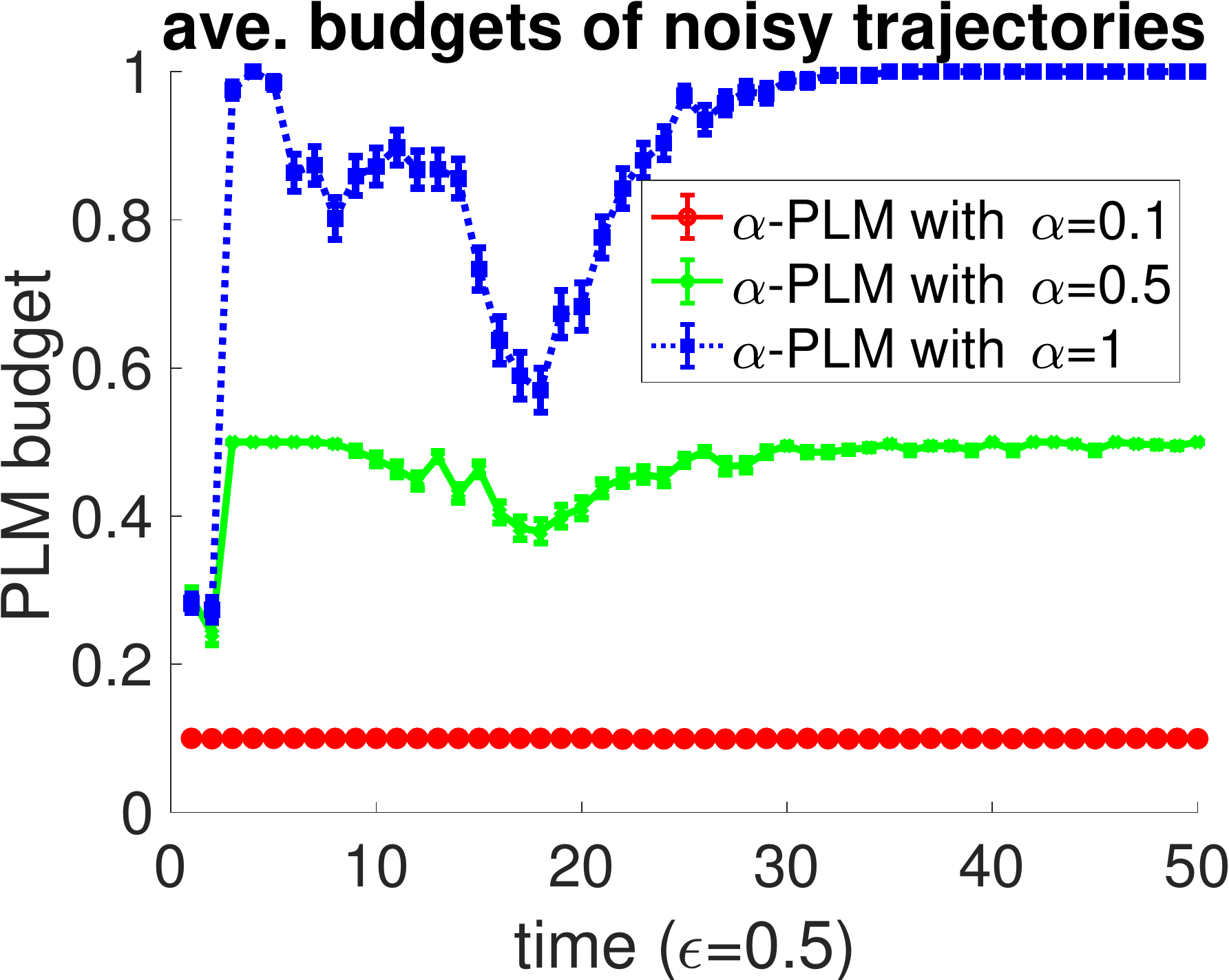}
		\vspace{-6pt}
\caption{{\scriptsize  Different PLMs for $\epsilon=0.5$.}}
	\label{Figure-expmt-B44}
\end{subfigure}
	\caption{{\footnotesize  $\textsc{Presence}(\mathcal{S}=[1:10],\mathcal{T}=[16:20])$ on synthetic data.}}
	\label{Figure-expmt-B33-44}
\end{figure}

\textbf{PriSTE with Geo-indistinguishability.}
 In Fig.\ref{Figure-expmt-B11-22}(a), it turns out that, 0.2-PLM  satisfies 1-spatiotemporal event privacy with only slight privacy budget reduction, and  satisfies 0.5-spatiotemporal event privacy with few budget reduction, but need to reduce more privacy budgets (to be stricter) in order to achieve 0.1-spatiotemporal event privacy.
 Similar results can be observed in Fig.\ref{Figure-expmt-B11-22}(b) and  Fig.\ref{Figure-expmt-B33-44}.
 We also observe that the standard deviation is larger for weaker LPPMs since these privacy budgets need to be frequently calibrated.
 Hence, we can conclude that a stricter PLM for location privacy can protect spatiotemporal event without much calibration, but a more loose PLM may need to reduce its privacy budget significantly for $\epsilon$-spatiotemporal event privacy.

Comparing Fig.\ref{Figure-expmt-B11-22} with  Fig.\ref{Figure-expmt-B33-44}, where the events are defined on time periods 4$ \sim $8 and 16$ \sim $20 respectively,  we can see that privacy budgets  tend to be reduced during the defined time periods.
This indicates that the final $ \alpha $ used by PLM at each timestamp may disclose the definition of spatiotemporal event as we discussed in Section \ref{sec:case_study1}.
Hence, a local model is needed for PriSTE framework.

\begin{figure}[!htbp]
	\centering
	\begin{subfigure}{0.24\textwidth}
		\centering
		\includegraphics[width=4.0cm]{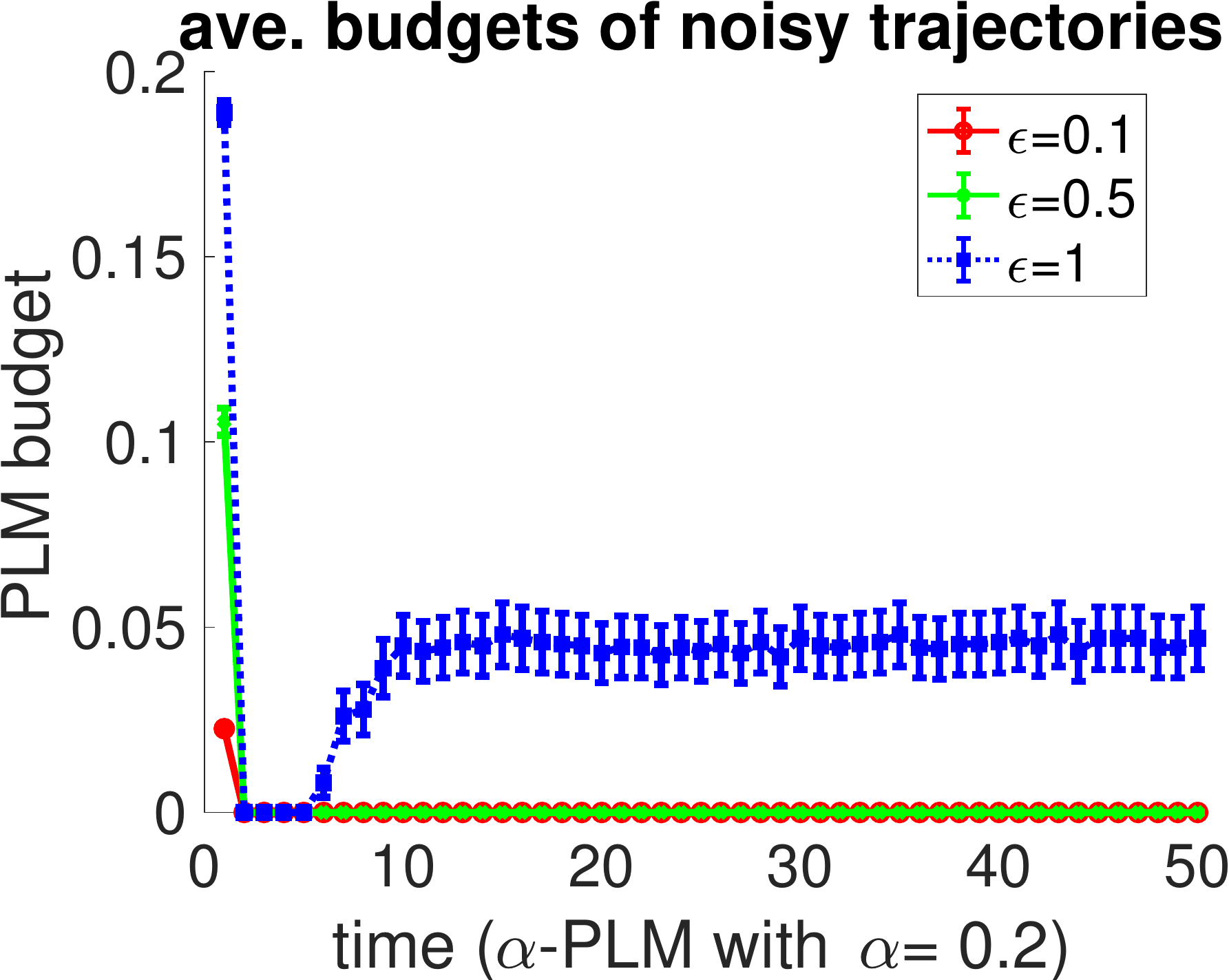}
		\caption{{\footnotesize $ 0.2 $-PLM  for different  $\epsilon$.}}
		\label{Figure-expmt-51}
	\end{subfigure}
	\begin{subfigure}{0.24\textwidth}
		\centering
		\includegraphics[width=4.0cm]{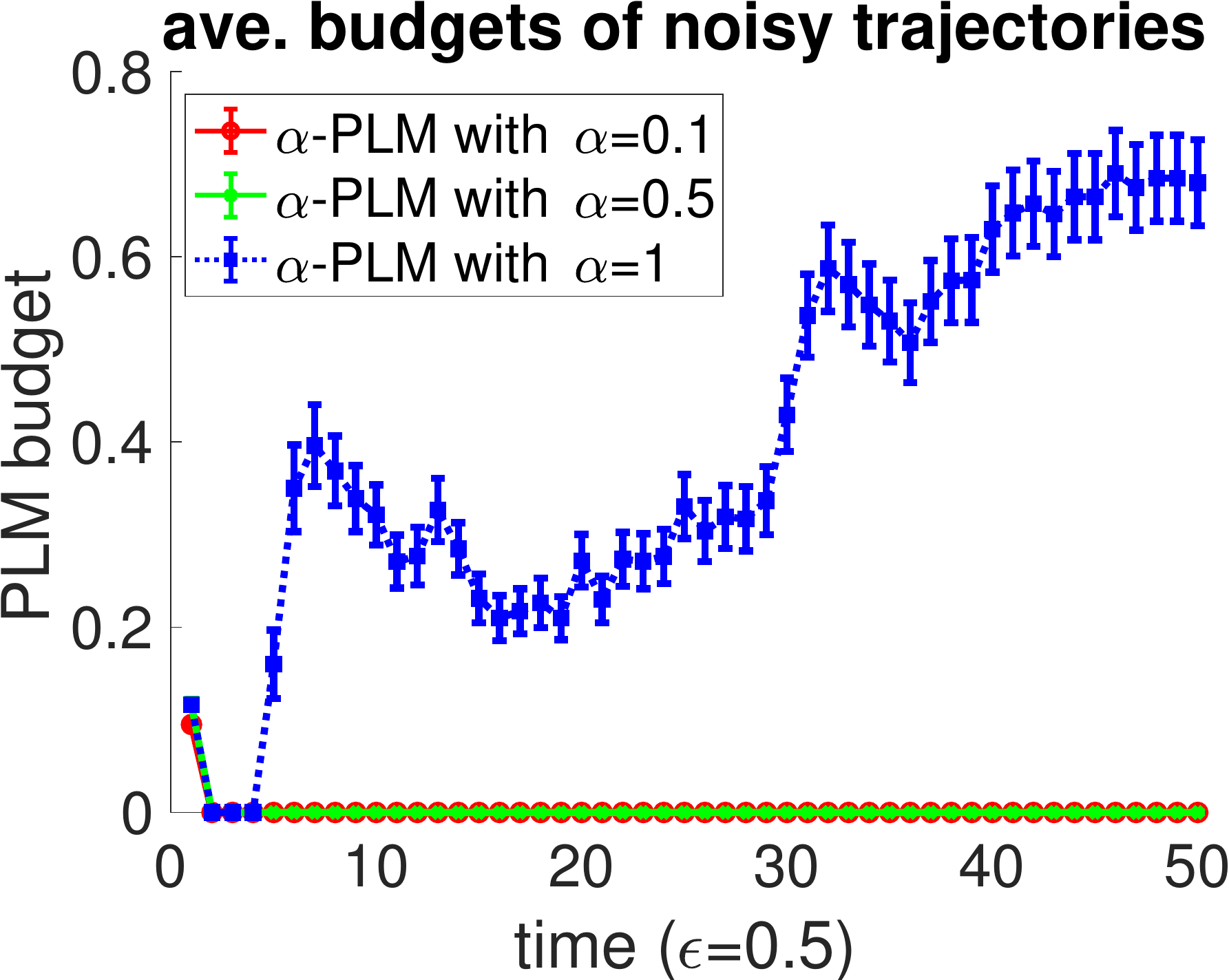}
		\caption{{\footnotesize Different PLMs for $\epsilon=0.5$.}}
		\label{Figure-expmt-B52}
	\end{subfigure}
	\caption{{\footnotesize  Protecting two events $\textsc{Presence}(\mathcal{S}=[1:10],\mathcal{T}=[4:8])$ and $\textsc{Presence}(\mathcal{S}=[1:10],\mathcal{T}=[16:20])$ on synthetic data.}}
	\label{Figure-expmt-B5}
\end{figure}

%
\textbf{Protecting multiple events.}
Fig.\ref{Figure-expmt-B5} depicts  the  utilities when protecting two events sequentially using Algorithm \ref{alg-priv-check1}.
We can see that the utility is much worse than protecting each single event  in Fig.\ref{Figure-expmt-B11-22} or  Fig.\ref{Figure-expmt-B33-44}  because the algorithm needs to simultaneously check if $\epsilon$-spatiotemporal event privacy is satisfied for \textit{both} events at each time. 
If no perturbed location satisfying the privacy requirement of  both events simultaneously, the algorithm needs to halve the privacy budget  until finding an appropriate output.

\begin{figure}[h]
	\centering
	\begin{subfigure}{0.24\textwidth}
		\centering
		\includegraphics[width=4.0cm]{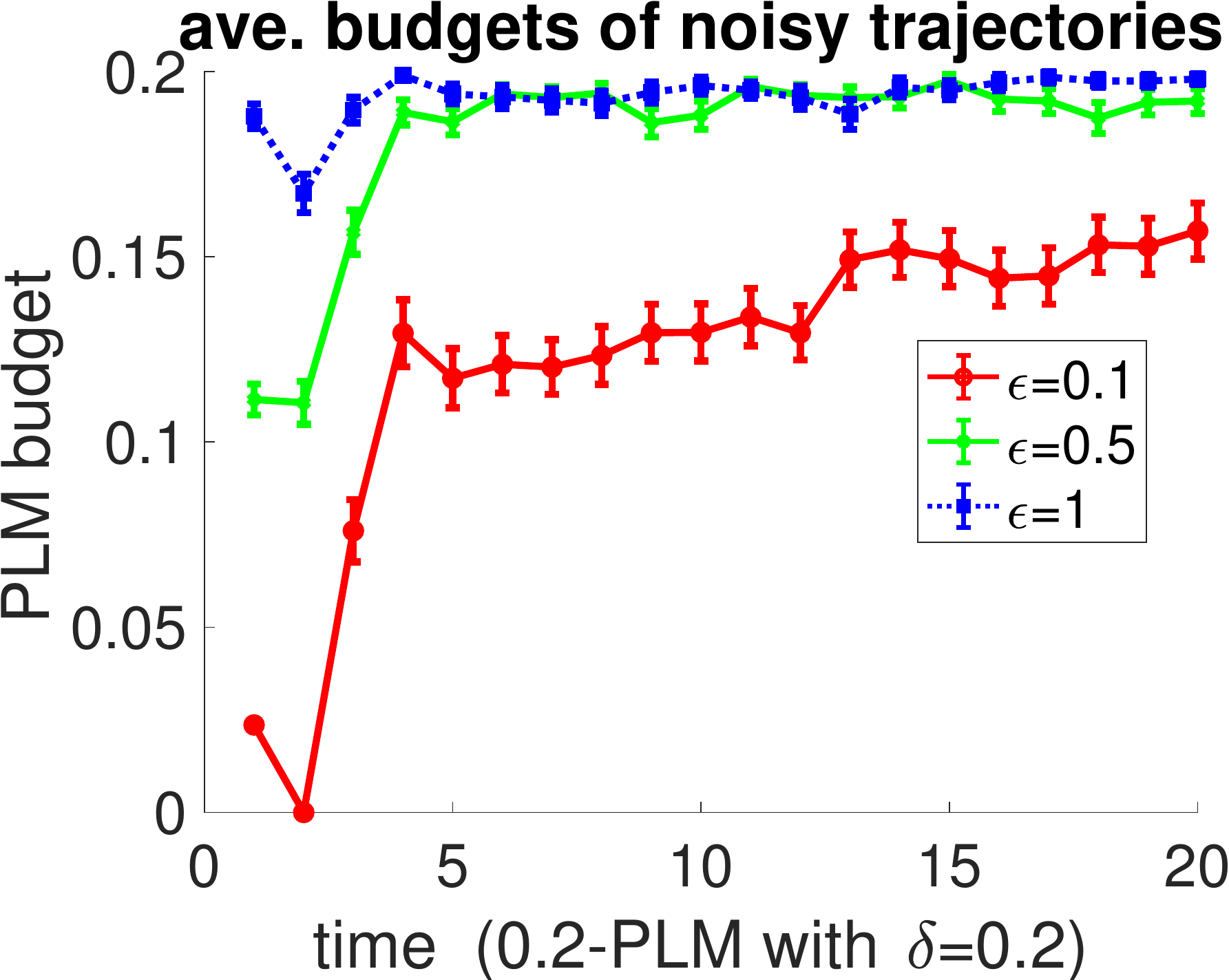}
		\vspace{-6pt}
		\caption{{\footnotesize $ 0.2 $-PLM with $ \delta $-location set privacy  for different  $\epsilon$.}}
		\label{Figure-expmt-61}
	\end{subfigure}
	\begin{subfigure}{0.24\textwidth}
		\centering
		\includegraphics[width=4.0cm]{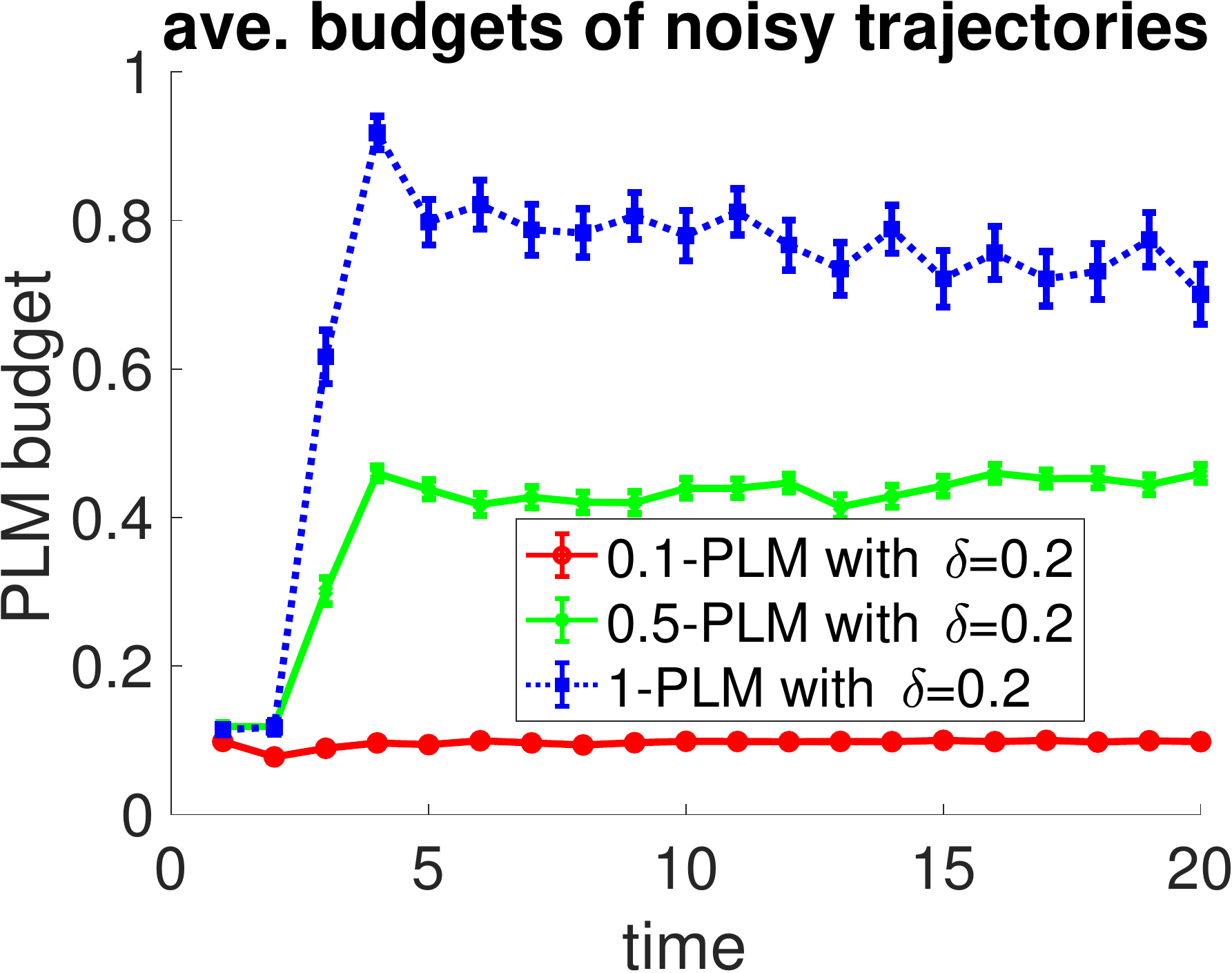}
		\vspace{-6pt}
		\caption{{\footnotesize  Different PLMs with $ \delta $-location set privacy for $\epsilon=0.5$.}}
		\label{Figure-expmt-B62}
	\end{subfigure}
	\caption{{\footnotesize  $\textsc{Presence}(\mathcal{S}=[1:10],\mathcal{T}=[4:8])$ on synthetic data.}}
	\label{Figure-expmt-B6}
\end{figure}

\textbf{PriSTE with $\delta$-Location Set Privacy.}
In Fig.\ref{Figure-expmt-B6}, we show the utility of PriSTE with LPPMs that satisfy $\delta$-Location Set Privacy (Algorithm \ref{alg-priv-check2}).
Comparing  Fig.\ref{Figure-expmt-B6} with Fig.\ref{Figure-expmt-B11-22}, although both of them are using 0.2-PLM, the essential difference between them is the privacy metric: the former satisfies $\delta$-location set privacy and the latter satisfies geo-indistinguishability, i.e., 0.2-PLM in  Fig.\ref{Figure-expmt-B6} has a constrained output domain.
We can see that such  $0.2$-PLM in Fig.\ref{Figure-expmt-B6} has to reduce more privacy budgets to achieve $\epsilon$-spatiotemporal event privacy. 
Intuitively, it is because the privacy metric of $\delta$-location set privacy  implies a weaker privacy guarantee  and its LPPM has to be stricter (using a smaller privacy budget) for protecting the event.

\vspace{-15pt}

\subsection{ Utility over Timestamps }
\label{sec:utility2}
\begin{figure}[h]
 	\centering
 	\begin{subfigure}{0.24\textwidth}
 		\centering
 		\includegraphics[width=4.0cm]{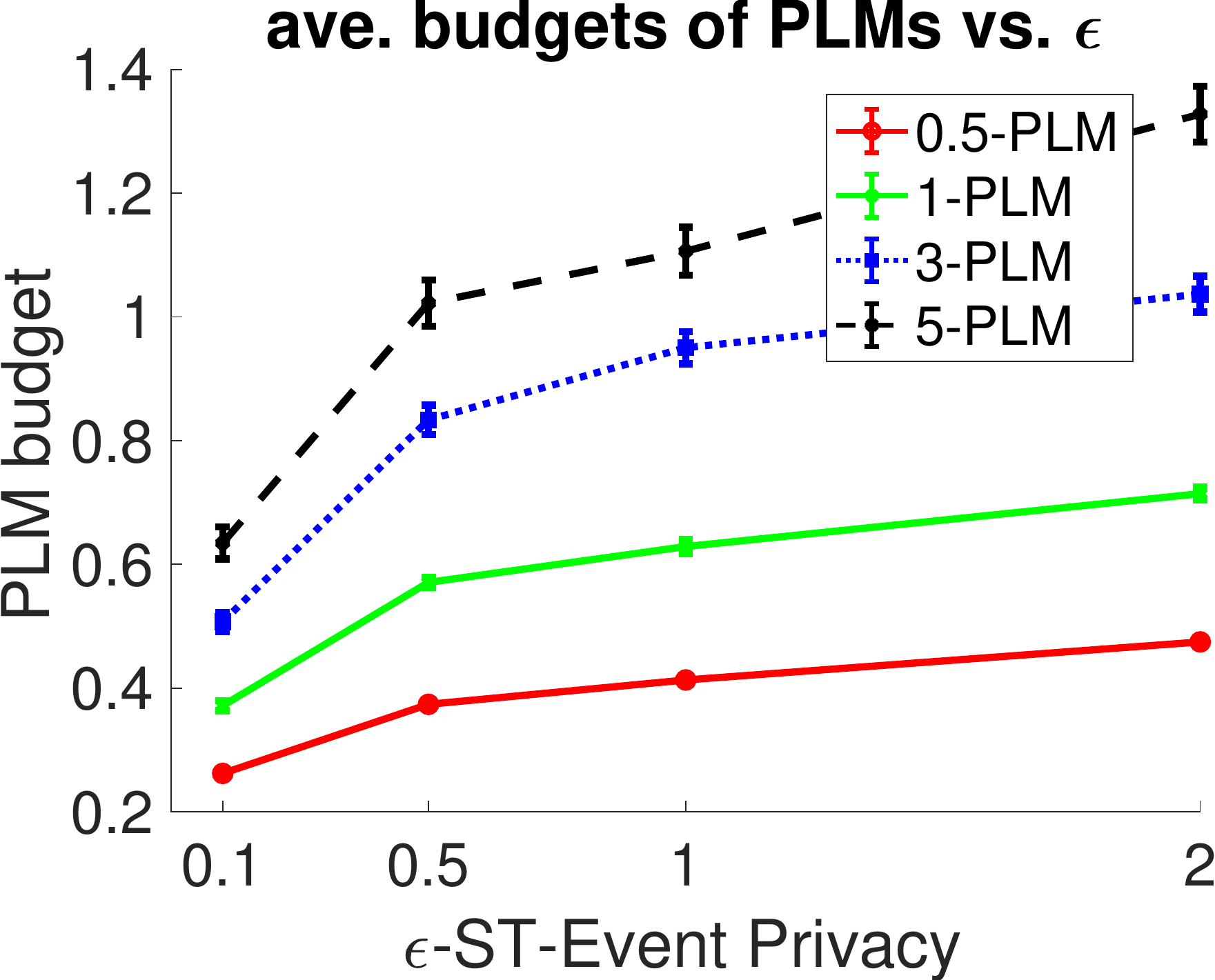}
 	\end{subfigure}
 	\begin{subfigure}{0.24\textwidth}
 		\centering
 		\includegraphics[width=4.0cm]{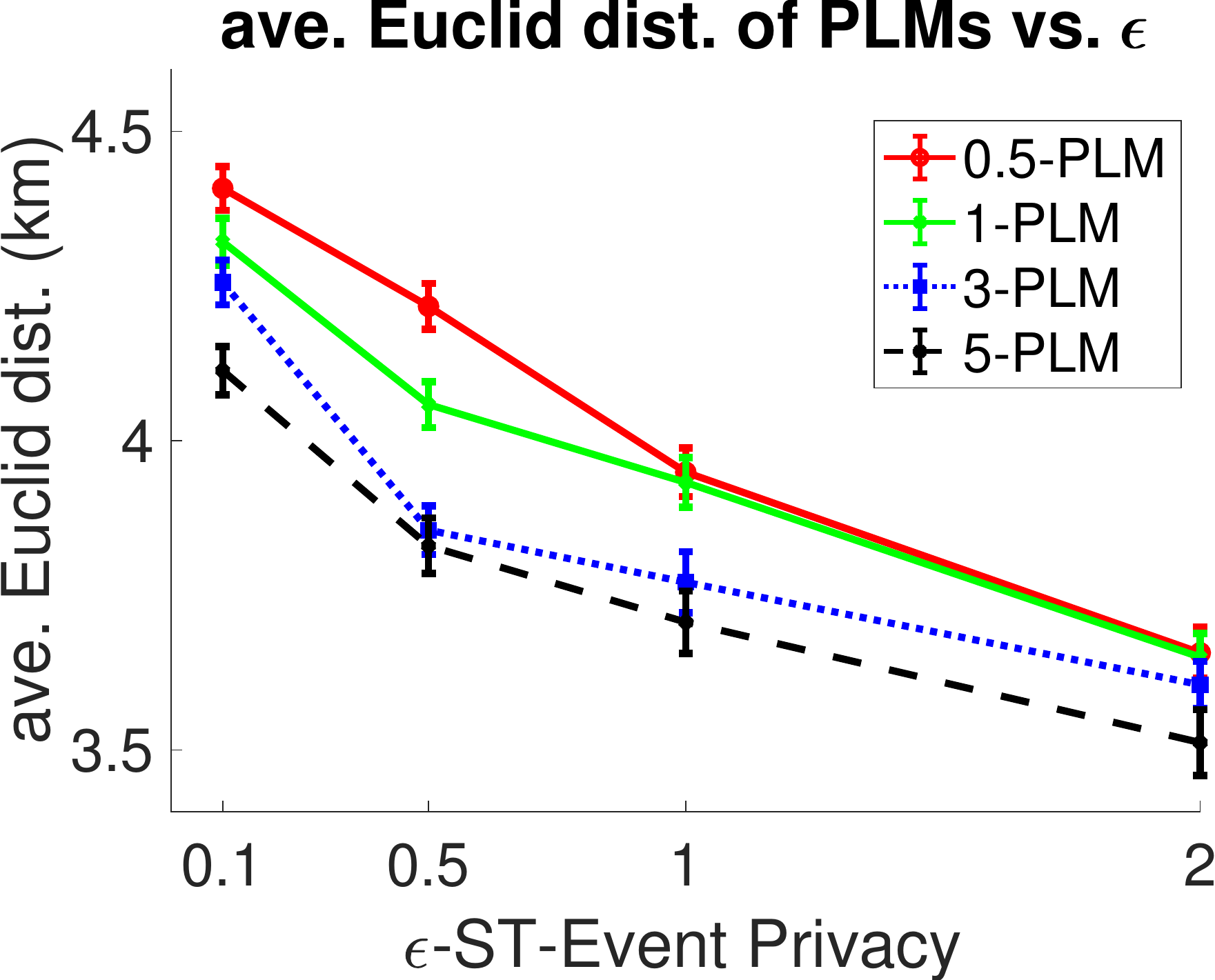}
 	\end{subfigure}
 	\caption{{\footnotesize  $\textsc{Presence}(\mathcal{S}=[1:10],\mathcal{T}=[4:8])$ on Geolife data.}}
 	\label{Figure_C1_12}
 \end{figure}

\begin{figure}[!htbp]
	\centering
	\begin{subfigure}{0.24\textwidth}
		\centering
		\includegraphics[width=4cm]{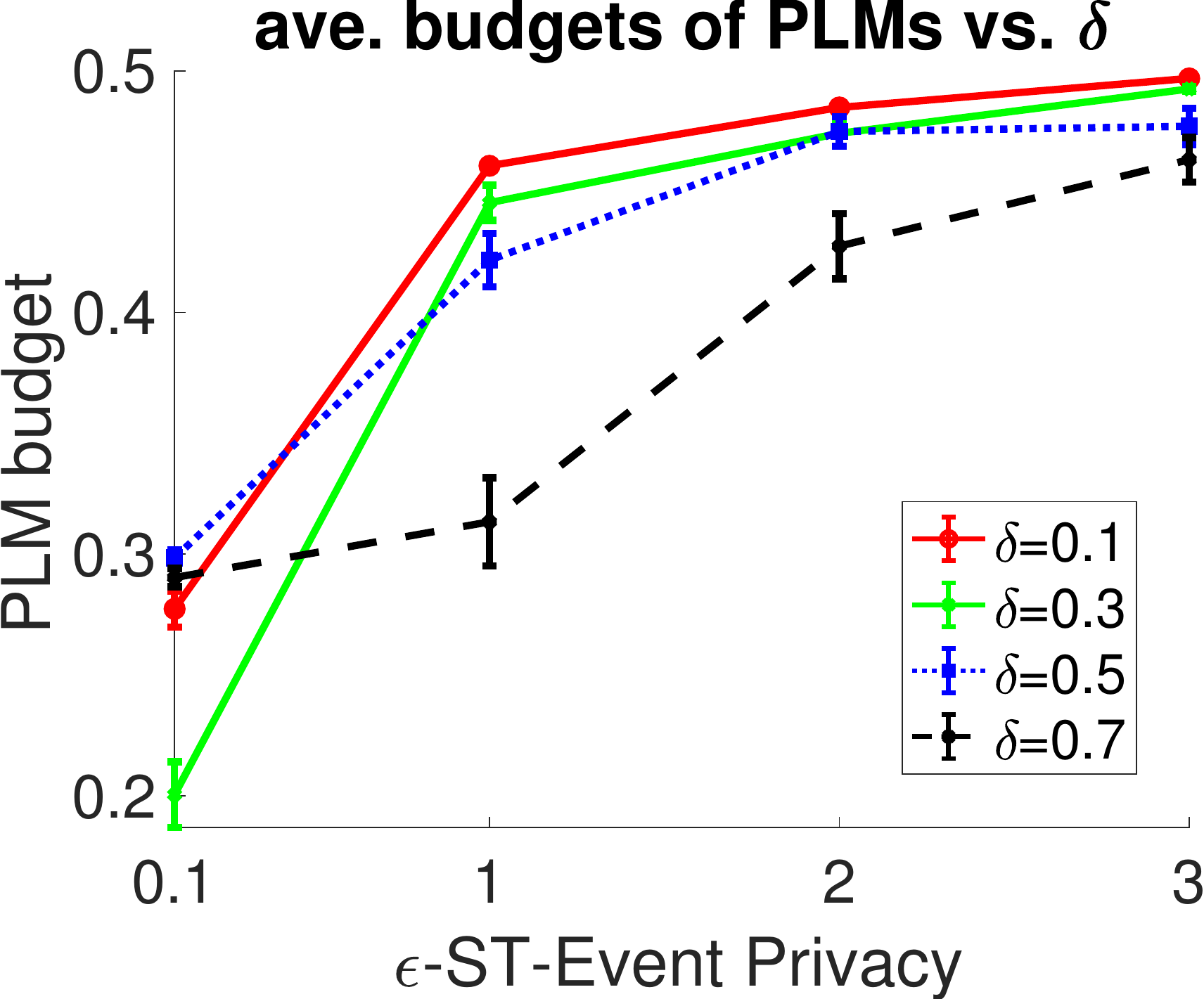}
		\label{Figure-expmt-C2-1}
	\end{subfigure}
	\begin{subfigure}{0.24\textwidth}
		\centering
		\includegraphics[width=4cm]{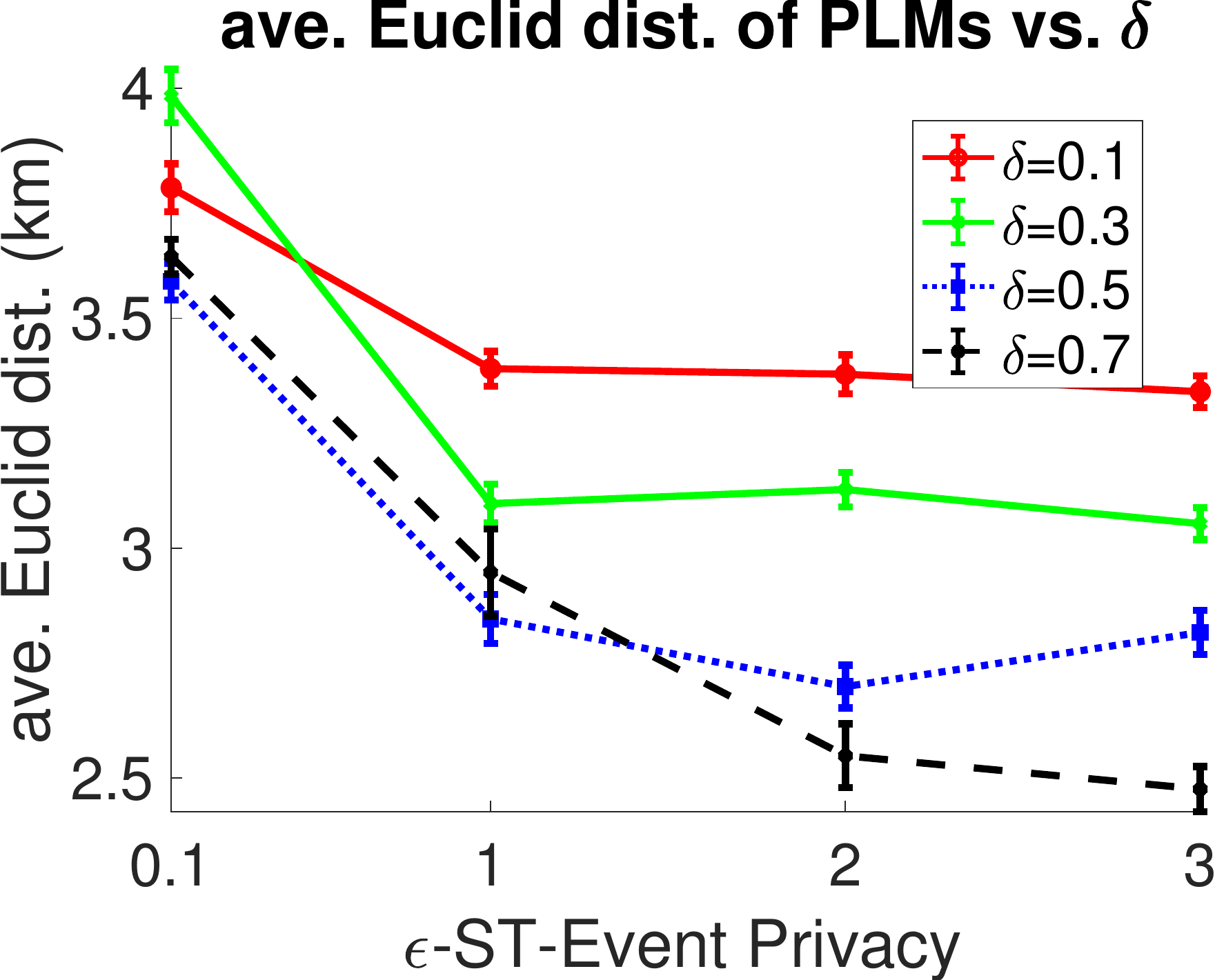}
		\label{Figure-expmt-C2-2}
	\end{subfigure}
	\caption{{\footnotesize  $\textsc{Presence}(\mathcal{S}=[1:10],\mathcal{T}=[4:8])$ on Geolife data (0.5-PLM with $ \delta $-location set privacy).}}
	\label{Figure_C2_12}
\end{figure}

In this section,  we demonstrate the utility against different factors on the Geolife data and synthetic data.
Figures \ref{Figure_C1_12} and \ref{Figure_C2_12} are for protecting \textsc{Presence} event.
Due to space limitation, the  results of protecting  \textsc{Pattern} event are included in Appendices.
Different from the utility in previous section which is averaged at each time, this section displays  the utility  that is further averaged over timestamps.
Hence, in the left parts of Figures \ref{Figure_C1_12} and \ref{Figure_C2_12} (ave. budget), the steeper lines indicate the budget may be reduced heavily at some timestamps.
 Generally, the utility increases with a larger $\epsilon$ in Fig.\ref{Figure_C1_12} and Fig.\ref{Figure_C2_12}.
 
 \textbf{Utility vs. $\alpha$-geo-indistinguishability.}
 In Fig.\ref{Figure_C1_12}, we can see that a larger $\alpha$-PLM needs to be calibrated heavily (i.e., steeper) for a small $\epsilon$.
 Interestingly,   PLMs with larger average budgets (in the left figures) may not necessarily have better utility in terms of Euclidean distance.
 For example,  at $ \epsilon=0.5$, the  Euclidean distance of  5-PLM and 3-PLM are very close; at $\epsilon=1 $ or $ 2 $,   0.5-PLM and 1-PLM appear to have almost  the same Euclidean distance.
 The reason is that PLMs that have  larger \textit{average}  budgets may have extremely small budgets at some timestamps, which results in the higher  \textit{average} Euclidean distance  over timestamps.


 \textbf{Utility vs. $\delta$-location set privacy.}
In Fig.\ref{Figure_C2_12}, we can see that 	 a PLM with  a larger $\delta$ tends to have a smaller average budget.
It is because the PLM with a larger $\delta$ indicates  a weaker privacy metric.
Hence, the PLM needs to be stricter (i.e, a small budget) to achieve spatiotemporal event privacy.
However, such PLM may have a better utility in terms of Euclidean distance as shown in right figure of Fig.\ref{Figure_C2_12}.
The reason is that  $\delta$-location set privacy with a larger $\delta$ restricts the output domain significantly, which makes perturbed location close to the true location with a high probability.
The results are in line with  the main purpose of $\delta$-location set privacy:  to have a better trade off between utility and privacy.

%
%

\begin{figure}[h]
	\centering
	\begin{subfigure}{0.24\textwidth}
		\centering
		\includegraphics[width=4cm]{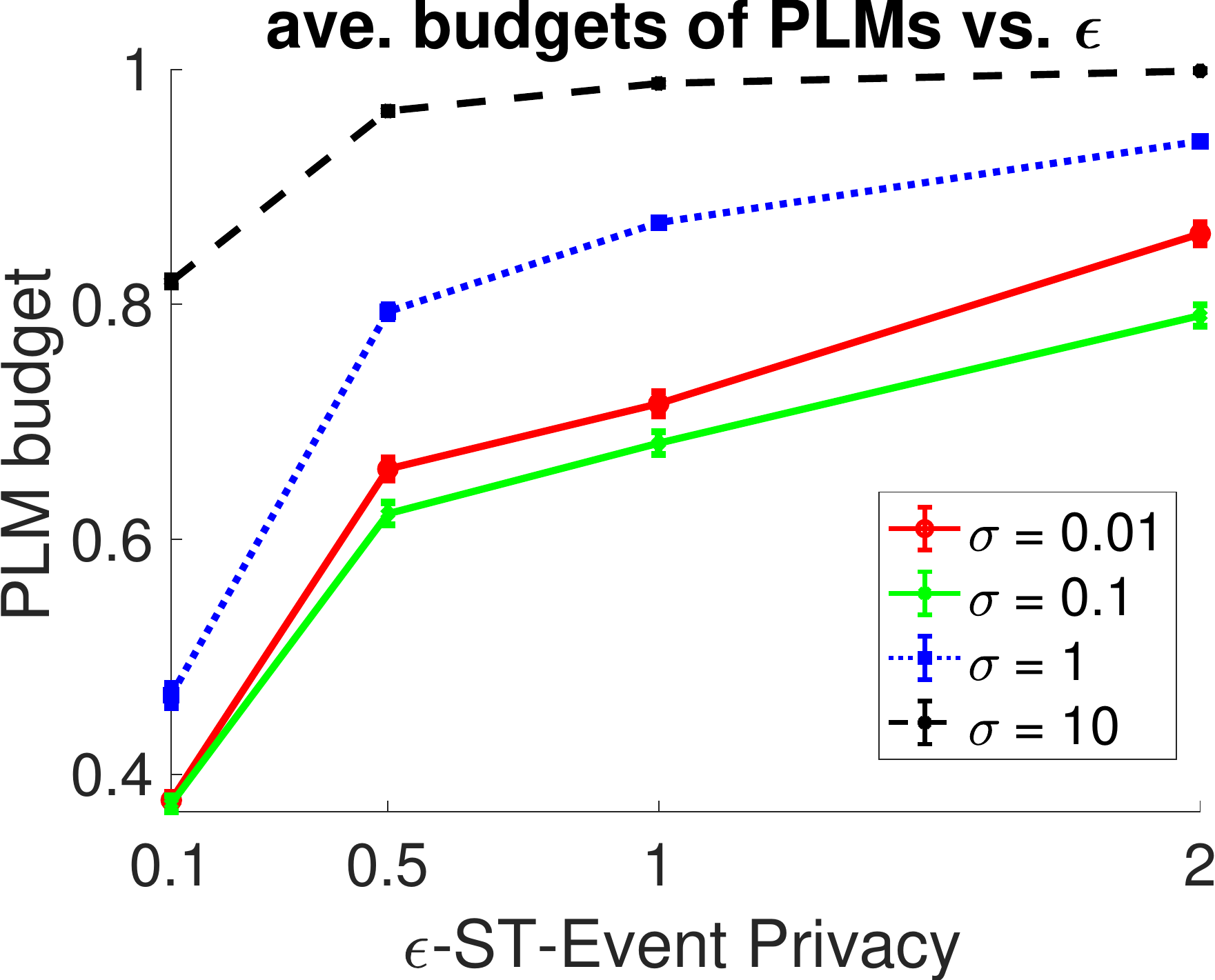}
		\label{Figure_C4_2}
	\end{subfigure}
	\begin{subfigure}{0.24\textwidth}
		\centering
		\includegraphics[width=4cm]{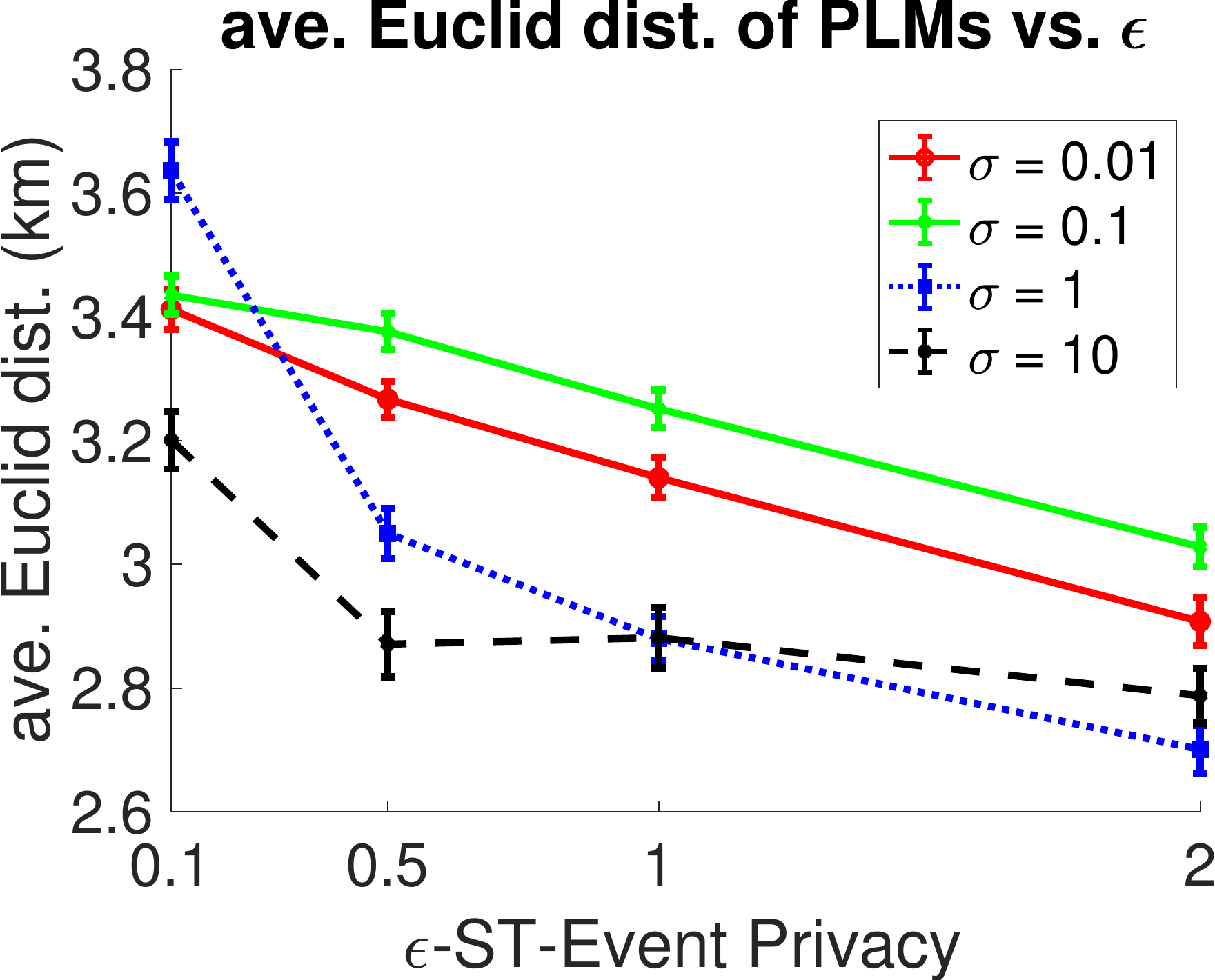}
		\label{Figure_C4_1}
	\end{subfigure}
	\caption{{\footnotesize  $\textsc{Presence}(\mathcal{S}=[1:10],\mathcal{T}=[4:8])$ on synthetic data (1-PLM with geo-indistinguishability).}}
\label{Figure_C4_12}
\end{figure}

 \textbf{Utility vs. Transition Matrices.}
 We compare the utility against  transition matrices that have different strength of mobility patterns.
As we explained previously, a smaller $\sigma$ indicates a more significant mobility pattern.
Fig.\ref{Figure_C4_12} shows that, for the same LPPM, it is hard to protect a spatiotemporal event if user's mobility pattern is significant, i.e., the LPPM needs to be very strict by using a small privacy budget.
We also observe that there is no best LPPM for all $\epsilon$-spatiotemporal event privacy in terms of Euclidean distance.
 	
\vspace{-10pt}
\subsection{Runtime}
\label{sec:rt}

We name the size of $ \mathcal{T} $ and the size of $ \mathcal{S} $ as  \textit{event length} and \textit{event width}, respectively.
We also report the performance evaluation on \textit{conservative release} described  in Section \ref{sec:case_study1}.

\begin{figure}[h]
	\centering
	\includegraphics[width=8.8cm]{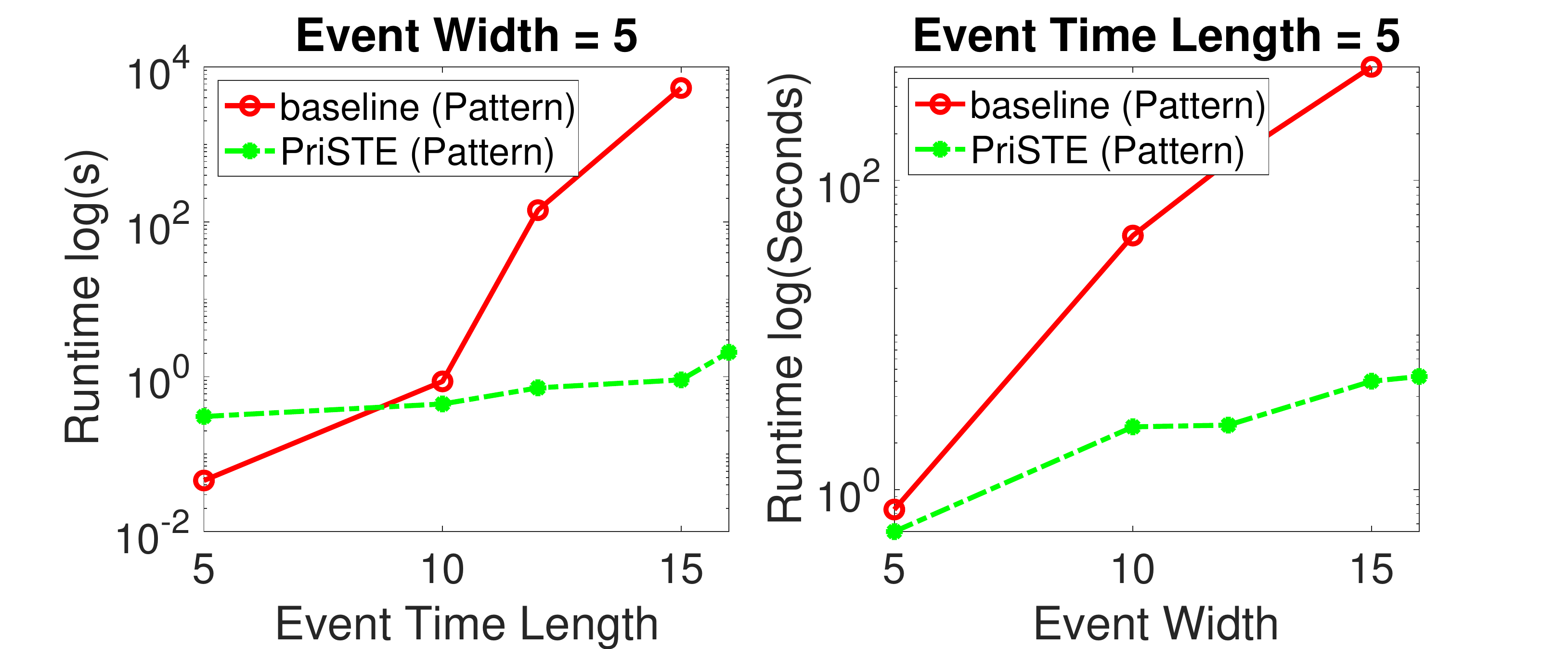}
	\caption{{\small Runtime evaluation.}}
	\label{Figure-expmt-runtime}
\end{figure}

\noindent\textbf{Runtime vs. Event Length}.
We fix the event width as 5 and test 100 random events with length ranging from 5 to 15.
Fig.\ref{Figure-expmt-runtime} shows that the average runtime of the baseline  is exponential to event length and the runtime of our method  is linear to  the event length.

\noindent\textbf{Runtime vs. Event Width}.
We fix the event length as 5 and test 100 random events with width ranging from 5 to 15.
Fig.\ref{Figure-expmt-runtime} shows that the average runtime of the baseline  is exponential to the event width, while our method is polynomial to the event width, which is in line with the  complexity of {\small $ O(m^3)$}.

\noindent\textbf{Runtime vs. Conservative Release}.
In Line 16 of Algorithm \ref{alg-priv-check1}, we set a threshold runtime in solving the quadratic program.
We do not release the perturbed location unless we are sure that Eq.\eqref{eqn-DP-check1} and Eq.\eqref{eqn-DP-check2} are true.
The threshold is a trade-off between runtime and utility as shown in Table \ref{tbl_runtime} among 100 runs.
We note that each runtime in Table  \ref{tbl_runtime}  includes the whole process of Algorithm \ref{alg-priv-check1}.
In our implementation, we set the threshold to 1 second.
We can see as the threshold increases, the number of conservative releases decreases, which results in increasing runtime.
On the other hand, the calibrated privacy budgets increasse as the threshold increases. 
This verifies the tradeoff between runtime and utility that can be achieved by the conservative release.

\begin{table}[h]
	\centering
	\scriptsize
	\begin{tabular}{|p{35pt}|p{35pt}|p{35pt}|p{35pt}|p{35pt}|}
		\hline
			\textbf{threshold (s)} & \textbf{ave. total runtime (s)} & $\# $ \textbf{of}  \space\space \textbf{Conservative} \textbf{Release} &  \textbf{ave.}  \textbf{privacy} 
			\textbf{budget}& \textbf{ave.}  \textbf{Euclidean} \textbf{dist. (km)}
		 \\\hline
		0.01 & 1.1 & \textbf{33} &  0.16 & 2.22 \\\hline
		0.1 & 2.6 & 30 &  0.23 & 1.51 \\\hline
			1 & 5.9 & 21 &  0.22 & 1.52 \\\hline
		2 & 10.4 & 12 &  {0.29} & \textbf{0.93 }\\\hline
		5 & {19.5} & 8 &  0.27 & 1.41 \\\hline
		none  & \textbf{52.5} & 0 &  \textbf{0.31} & 0.97 \\\hline
	\end{tabular}
	\caption{Runtime vs. threshold.}
	\label{tbl_runtime}
\end{table}


\section{Related Works}

\subsection{Location Privacy Preserving Mechanisms}

The LPPMs \cite{andres_geo-indistinguishability:_2013}\cite{xiao_protecting_2015}\cite{ghinita_preventing_2009}\cite{ardagna_obfuscation-based_2011}\cite{hwang_novel_2012} \cite{takagi_geo_2019} generally use some
obfuscation methods, like spatial cloaking, cell merging, location precision reduction or dummy cells, to manipulate the probability distribution of
users' locations. 
As differential privacy becomes a standard for privacy protection, \cite{andres_geo-indistinguishability:_2013} proposed a Geo-indistinguishability notion based on differential privacy and a planar Laplace mechanism to achieve it. Xiao et al.  \cite{xiao_protecting_2015}\cite{xiao_loclok:_2017} studied how to protect location privacy under temporal correlations with an optimal differentially private mechanism. 
Rodriguez-Carrion et al. \cite{rodriguez-carrion_entropy-based_2015}  also studied the effect of temporal dependencies on entropy-based location privacy metric.
They proposed a new privacy metric \textit{entropy rate} and perturbative mechanisms based on it, which can be an alternative LPPM in our framework for protecting spatiotemporal event privacy.
Several studies \cite{shokri_quantifying_2011} \cite{bordenabe_optimal_2014} \cite{shokri_privacy_2015} \cite{xiaolan-2019}  \cite{cao_differentially_2015} \cite{cao_differentially_2016}  tried to achieve an optimal trade-off between the utility of applications and the privacy guarantee in the LPPMs.
Overall, above works all focused on the mechanisms of location privacy, which can be used in our framework as given LPPMs. Whereas we study a new problem of spatiotemporal event privacy. 

\subsection{Inferences on Location}
Various inference attacks can be carried out based on location information and external information such as moving patterns. In the aggregated setting, recent works have studied location or trajectory recovery attacks from aggregated location data\cite{liu_location_2018}  \cite{xu_trajectory_2017}  
 or proximity query results from location data \cite{argyros_evaluating_2017}.
We mainly discuss the individual setting that is closely related to our work.
Studies
\cite{shokri_quantifying_2011}\cite{cao_quantifying_2017}\cite{cao_quantifying_2019} \cite{cao_contpl:_2018} investigated the question of how to formally quantify the privacy of existing LPPMs, given an adversary who can model users' mobility using a Markov process learned from population.  
Liao et al. \cite{liao_learning_2004} used a hierarchical Markov model to learn and infer a user's trajectory based on the places and temporal patterns they visited.  
Qiao et al. \cite{qiao_putmode:_2010} used the Continuous Time Bayesian Networks to predict uncertain trajectories of moving objects.
Li et al. \cite{li_swarm:_2010} uses frequent mining approach to find  moving objects that move within arbitrary shape of clusters for certain timestamps that are possibly nonconsecutive.



\section{Conclusion and Future Work}
In this paper, we investigate a new type of pivacy goal: protecting spatiotemporal event, which has not been studied in literature.
We  formally define spatiotemporal events and design a privacy metric extending the notion of differential privacy.
We proposed PriSTE, a framework integrating an LPPM for protecting the spatiotemporal event privacy.
 An interesting direction is to find optimal way for achieving both location privacy and spatiotemporal event privacy.
 Another question is how we can design a generic mechanism  for spatiotemporal event privacy.



\vspace{-1pt}

\bibliographystyle{IEEEtran}
\bibliography{ref/stevent}

\newpage
\clearpage

\begin{appendices}
	
\section{Proofs}

\subsection{Proof of Lemma \ref{lemma-post-before}}
\begin{proof}
	At timestamps $1$, {\footnotesize $\boldsymbol\alpha_1=[\boldsymbol\pi, \textbf{0}]\circ\tilde{\textbf{p}}_{o_1}$}. 
	For $t>1$, 
	{\footnotesize $\boldsymbol\alpha_t=\boldsymbol\alpha_1\textbf{M}_{1}\tilde{\textbf{p}}_{o_2}^\textbf{D}\cdots\textbf{M}_{t-1}\tilde{\textbf{p}}_{o_t}^\textbf{D}	=[\boldsymbol\pi,\textbf{0}]\left(  \tilde{\textbf{p}}_{o_1}^\textbf{D}\prod_{i=2}^{t}(\textbf{M}_{i-1}\tilde{\textbf{p}}_{o_i}^\textbf{D})\right)$}. By Lemma \ref{theo-prior}, 
	{\footnotesize $\Pr(\textsc{Event},o_1,o_2,\cdots,o_t)=\boldsymbol\alpha_t\prod_{i=t}^{end-1}\textbf{M}_i[\textbf{0},\textbf{1}]^\intercal$}. 
	Then Equation (\ref{eqn-post-before}) can be derived. 
\end{proof}

\subsection{Proof of Lemma \ref{lemma-post-after}}
\begin{proof}
	{\small $\Pr(o_1,o_2,\cdots,o_t,\textsc{Event}\textrm{ is true})=\sum_{k} \Pr \big(l_{end}=k,o_1,o_2,$}  {\small$\cdots,o_t, \textsc{Event}\textrm{ is true} \big)$}.
	By forward-backward algorithm, {\small $\Pr(l_t=s_{k},o_1,o_2,\cdots,o_t)=\alpha_{t}^{k}\beta_{t}^{k}$}. Thus we only need to derive the $\boldsymbol\alpha_{end}$ and $\boldsymbol\beta_{end}$ and compute the sum of {\small $\boldsymbol\alpha_{end}\circ \boldsymbol\beta_{end}$} in the world where the $\textsc{Event}$ is true.  
	By Lemma \ref{lemma-post-before}, 
	{\small $\boldsymbol\alpha_{end}=[\boldsymbol\pi,\textbf{0}]\left(  \tilde{\textbf{p}}_{o_1}^\textbf{D}\prod_{i=2}^{end}(\textbf{M}_{i-1}\tilde{\textbf{p}}_{o_i}^\textbf{D})\right)$}. 
	By backward algorithm, {\small $\boldsymbol\beta_{end}
		=[\textbf{1},\textbf{1} ]\prod_{i=t-1}^{end}(\tilde{\textbf{p}}_{o_{i+1}}^\textbf{D}\textbf{M}_i^\intercal)$}
	with {\small $\boldsymbol\beta_t=[\textbf{1},\textbf{1} ]$}. 
	Thus {\small $\Pr(\textsc{Event},o_1,o_2,\cdots,o_t)=(\boldsymbol\alpha_{end}\circ\boldsymbol\beta_{end})[\textbf{0},\textbf{1}]^\intercal
		=\boldsymbol\alpha_{end}(\boldsymbol\beta_{end} \circ [\textbf{0},\textbf{1}])^\intercal$}, which is equal to Equation (\ref{eqn-post-after}).
\end{proof}

\subsection{Proof of Theorem \ref{theo-eps-delta-DP}}

\begin{proof}[Proof Sketch]
	By Definition \ref{def:e-STE}, it is equivalent to prove {\small $f_{1}(\boldsymbol\pi)\leq 0$} and {\small $f_{2}(\boldsymbol\pi)\leq 0$} where 
	{\small $f_{1}(\boldsymbol\pi)=\frac{\Pr(o_{1},o_{2},\cdots,o_{T}, \textsc{Event})}{\Pr(\textsc{Event})}-e^{\epsilon}\frac{\Pr(o_{1},o_{2},\cdots,o_{T}, \lnot\textsc{Event})}{\Pr(\lnot\textsc{Event})}$} 
	and 
	{\small $f_{2}(\boldsymbol\pi)=\frac{\Pr(o_{1},o_{2},\cdots,o_{T}, \lnot\textsc{Event})}{\Pr(\lnot\textsc{Event})}-e^{\epsilon}\frac{\Pr(o_{1},o_{2},\cdots,o_{T}, \textsc{Event})}{\Pr(\textsc{Event})}$}. By Lemma \ref{theo-prior} and \ref{lemma-post-after}, Theorem \ref{theo-eps-delta-DP} can be derived. 
\end{proof}

\section{Naive Solutions}

\subsection{Computing Prior Probability of an Event}

Considering an event is  a set of Boolean expression of (location, time) predicates combined with AND and OR, a naive approach would be to enumerate all possible cases for the event and sum (correspond to OR) the product (correspond to AND) of the probabilities of each location predicate and such an approach would require exponential computation time. 
Due to space limitation, we omit a detailed algorithm for this naive solution.
Instead, we show an example as below.
\begin{figure}[h]
	\centering
	\includegraphics[width=7cm]{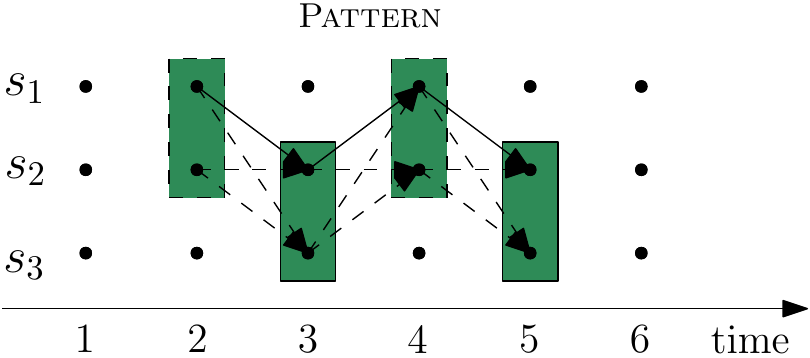}
	\caption{{\footnotesize $\Pr(\textsc{Pattern})=\sum Pr(\textrm{all trajectories rendering \textsc{Pattern}})$ for the $2^{4}$ trajectories satisfying the \textsc{Pattern} event defined above.}}
	\label{Figure-example-naive}
\end{figure}
\begin{example}
	\label{example-prior-pattern}
	In Fig.\ref{Figure-example-naive}, a \textsc{Pattern} event is defined by the shaded regions during timestamp $2$ to $5$. Thus $start=2$, $end=5$. The regions are {\small $\textbf{s}_2=[1,1,0]^{\intercal}$}, {\small $\textbf{s}_3=[0,1,1]^{\intercal}$}, {\small $\textbf{s}_4=[1,1,0]^{\intercal}$}, {\small $\textbf{s}_5=[0,1,1]^{\intercal}$}. To derive the probability of one trajectory, e.g. the solid lines in Fig.\ref{Figure-example-naive}, it would be 
	{\small $\Pr(l_2=s_{1})*
		\Pr(l_3=s_{2}|l_2=s_{1})*\Pr(l_4=s_{1}|l_3=s_{2},l_2=s_{1})*\Pr(l_5=s_{1}|l_4=s_{1},l_3=s_{2},l_2=s_{1})$}. Because there are $2^{4}$ trajectories for the \textsc{Pattern} event, {the prior probability of  \textsc{Pattern}, i.e.,} $ \Pr $(\textsc{Pattern}) is the sum of $2^{4}$ such probabilities.
\end{example}

\subsection{Computing Join Probability of an Event}

\begin{algorithm}[h]
	\scriptsize
	\caption{Naive Algorithm to Derive the Joint Probability of a \textsc{Pattern}}
	\begin{algorithmic}
		\Require{
			$\textbf{M}$, $\textbf{p}_{start-1}$: the probability at timestamp $start-1$, $\tilde{\textbf{p}}_{o_t}$, $\textsc{Pattern}$ in timestamp $start,\cdots, end$
		}
		\State{$p_{pattern}\gets 0$}
		\For{traj in \textsc{Pattern}}
		\Comment{exponential trajectories}
		\State{$\textbf{p}_{traj}\gets (\textbf{p}_{start-1}\textbf{M})\circ \tilde{\textbf{p}}_{o_t}$}
		\State{$p_{traj}=\textbf{p}_{traj}[l_{start}\ in\ traj]$}
		\For{$t$ in $\{start+1,\cdots,end\}$}
		\State{$s_{t-1}\gets l_{t-1}\ in\ traj$}
		\State{$s_{t}\gets l_t\ in\ traj$}
		\State{$p_{traj}\gets p_{traj}*m_{s_{t-1}s_{t}}*\tilde{\textbf{p}}_{o_t}[{s_{t}}]$}
		\EndFor
		\State{$p_{pattern}\gets p_{pattern}+p_{traj}$}
		\EndFor
	\end{algorithmic}
	\label{alg-naive}
\end{algorithm}

Let traj be a trajectory of the \textsc{Pattern}. There will be $|traj|$ trajectories of the \textsc{Pattern}. For example, in Figure \ref{Figure-example-naive}, the solid line is $traj=\{l_2=s_{1},l_3=s_{2},l_4=s_{1},l_5=s_{2}\}$, which means $l_2$ in traj is $s_{1}$, $\cdots$, $l_5$ in traj is $s_{2}$. Let $\textbf{M}$ be the transition matrix where $m_{ij}$ is the transition probability from state $i$ to state $j$. 
$\tilde{\textbf{p}}_{o_t}$ denotes the emission probability of observing $o_t$, 
$\tilde{\textbf{p}}_{o_t}=[
\Pr(o_t|l_t=s_1), \Pr(o_t|l_t=s_2),\cdots,\Pr(o_t|l_t=s_m)]$. Thus $\tilde{\textbf{p}}_{o_t}[s_{1}]=\Pr(o_t|l_t=s_1)$, $\cdots$, $\tilde{\textbf{p}}_{o_t}[s_{m}]=\Pr(o_t|l_t=s_m)$.
Let $\Pr(o_{start},\cdots,o_{end},traj)$ be the joint probability of the observations and the trajectory. Then $\Pr(traj | o_{start},\cdots,o_{end})=\frac{\Pr(o_{start},\cdots,o_{end},traj)}{\Pr(o_{start},\cdots,o_{end})}$. Thus we focus on the joint probability $\Pr(o_{start},\cdots,o_{end},traj)$.

\noindent{\bf Setup.} Let $start=2$, $\textbf{p}_{start-1}=\boldsymbol\pi$. For a given $\boldsymbol\pi$, Algorithm \ref{alg-naive} can be used to derive the joint probability of a \textsc{Pattern}. The runtime can be compared with the runtime of Equation (\ref{eqn-post-before}), which is also the joint probability. 

Note that to derive Equation (\ref{eqn-post-before}), we prefer ``vector*matrix'' instead of ``matrix*matrix'' for efficiency. Calculate it from left to right so that no matrix multiplication should be conducted.

\section{Examples}

\subsection{An Example of Prior Probability Computation}
We use an example to describe the  details in computation. 
\begin{example}[Prior Probability Computation]
	\label{example-running-2}
	Let us consider the computation of Example \ref{example-prior-presence}. For the $\textsc{Presence}$ event defined at $t=3$ and $t=4$, the transition matrix at $t=2$ and $t=3$ derived by Equation (\ref{new-M-1}) is the left matrix below; while the transition matrix at $t=1$ and $t\geq 4$ derived by Equation (\ref{new-M-2}) is the right matrix below. 
	\begin{myAlignSSS}
		\hspace{-1mm}
		\left[
		\begin{array}{cccccc}
			0 &0&0.7&0.1&0.2&0\\
			0&0&0.5 & 0.4&0.1&0\\
			0&0&0.9&0&0.1&0\\
			0&0&0&0.1&0.2&0.7\\
			0&0&0&0.4&0.1&0.5\\
			0&0&0&0&0.1&0.9
		\end{array}
		\right]
		\left[
		\begin{array}{cccccc}
			0.1 &0.2&0.7&0&0&0\\
			0.4&0.1&0.5 & 0&0&0\\
			0&0.1&0.9&0&0&0\\
			0&0&0&0.1&0.2&0.7\\
			0&0&0&0.4&0.1&0.5\\
			0&0&0&0&0.1&0.9
		\end{array}
		\right]
	\end{myAlignSSS}
	
	\vspace{-10pt}
	\begin{minipage}{0.48\textwidth}
		\begin{tabular}{p{23pt}p{110pt}p{105pt}}
			&$\textbf{M}_{2}, \textbf{M}_{3}$ & $\textbf{M}_{1}, \textbf{M}_{4}, \textbf{M}_{5}$
		\end{tabular}
	\end{minipage}
	
	Thus by Lemma \ref{theo-prior}, {\footnotesize $\Pr(\textsc{Presence})=[\boldsymbol\pi,\textbf{0}]\textbf{M}_{1}\textbf{M}_{2}\textbf{M}_{3}[\textbf{0},\textbf{1}]^{\intercal}$}, which is a function of $\boldsymbol\pi$ as {\footnotesize $\Pr(\textsc{Presence})=\boldsymbol\pi[0.28, 0.298,0.226]^{\intercal}$}.
\end{example}

\end{appendices}

\end{document}